\documentclass[twoside]{article}

\usepackage{authblk}

\usepackage[round]{natbib}
\usepackage{amsmath}
\usepackage[hyphens]{url}
\usepackage{hyperref}
\usepackage{todonotes}
\usepackage[margin=2.5cm]{geometry}
\usepackage{multirow}
\usepackage{textcomp}
\usepackage{fancyvrb}
\newcommand{\ignore}[1]{}

\providecommand{\tightlist}{%
  \setlength{\itemsep}{0pt}\setlength{\parskip}{0pt}}

\title{Inconsistency in Conference Peer Review: Revisiting the 2014 NeurIPS Experiment}
\author[$\star$]{Corinna Cortes}
\author[$\dagger$]{Neil D. Lawrence}
\affil[$\star$]{Google Research, New York}
\affil[$\dagger$]{Computer Lab, University of Cambridge}
\begin{document}

\maketitle

%

%

%

\begin{abstract}
    In this paper we revisit the 2014 NeurIPS experiment that examined
inconsistency in conference peer review. We determine that 50\% of the variation
in reviewer quality scores was subjective in origin. Further, with seven years
passing since the experiment we find that for \emph{accepted} papers,
there is no correlation between quality scores and impact of the paper
as measured as a function of citation count. We trace the fate of
rejected papers, recovering where these papers were eventually
published. For these papers we find a correlation between quality scores
and impact. We conclude that the reviewing process for the 2014
conference was good for identifying poor papers, but poor for
identifying good papers. We give some suggestions for improving the
reviewing process but also warn against removing the subjective element.
Finally, we suggest that the real conclusion of the experiment is that
the community should place less onus on the notion of `top-tier
conference publications' when assessing the quality of individual
researchers. 
\end{abstract}

\hypertarget{introduction}{%
\section{Introduction}\label{introduction}}

The NeurIPS conference is one of the premier conferences in machine
learning.  The papers presented at this conference have
proven to be highly influential including breakthrough papers in
supervised and unsupervised learning, structure prediction, etc. They provide theoretical, algorithmical and experimental justification
for the proposed techniques. These ideas have gone on to have widespread
societal impact.

Since around 2010, the increased interest in AI has
significantly increased the size of the conference and already in
2014, questions around the quality of the reviewing process in
connection with this increased load on reviewers had been raised.

In 2014 as Program Chairs of the NeurIPS conference, we implemented the NeurIPS experiment. The experiment
was designed to assess the consistency of the conference peer
reviewing process. From the conference 10\% of the papers were
randomly chosen to be reviewed by two independent program
committees. The objective was to determine if decision making was
consistent across these two committees.  The results showed that the
decisions between the two committees was better than random, but still
surprised the community by how low it was.  A particular focus of community concern was
the fact that the two committees were 
very inconsistent about which papers were selected to appear at the
conference, so that if the review process had been independently
rerun, about half the papers published at the conference would have
been different. This experiment is being repeated by the 2021 NeurIPS Program Chairs.

We revisit the 2014 conference data and explore these numbers further
in three ways. First, we use the fact that reviewer scores underwent a
calibration process during the conference. This process was focused on
eliminating bias in reviewer scale interpretation, but it also
quantifies the subjectivity of individual reviewer scores. Through a
simulation study we demonstrate that this subjectivity is at the
heart of the inconsistency. Second, we explore whether these scores
correlated with paper citation scores.  Taking citation scores as a
proxy for paper impact,\footnote{There are problems with using
  citation scores as a way of assessing impact, see
  e.g.\ \cite{Neylon-article09} for a discussion, but they have the
  advantage of being an objective, community driven measure. Seven
  years having passed since publication, and the papers have had a
  chance to establish themselves.} we collected citation counts for
each of the around 400 published papers from Semantic
Scholar.\footnote{\url{https://www.semanticscholar.org/}} We find no
correlation between paper quality scores and the paper's eventual
impact. Finally, we analyse rejected papers from the conference. We
searched Semantic Scholar for papers in the literature with similar titles by the same
lead author allowing us to track the final outlet
for 680 papers that were rejected by the 2014 NeurIPS conference, as
well as their associated citation counts. For these papers we do find
correlation between quality scores and citation counts.

From these analyses we conclude that inconsistency in the conference
reviewing process is a consequence of the subjectivity in reviewer
assessments. And that in the high-scoring range, reviewer quality
scores are not a good proxy for citation impact. However, reviewers
seem better at identifying weaker papers: low-scoring papers resulted (on
average) in lower impact papers. In our conclusions, we argue that the different
facets of a paper could be better assessed with clearer scoring
criteria. This would give program chairs more flexibility in guiding
the nature of the conference.

Before discussing our experiments, we will start with a brief
reminder of the NeurIPS experiment.

\section{The NeurIPS Experiment}
\label{review-of-the-conference-and-the-experiment}

The NeurIPS 2014 conference was held in Montreal with 2,581 attendees at the
conference, associated workshops, and tutorials. At NeurIPS 2014, each paper was assigned to an Area Chair and at least
three reviewers. Final decisions about papers were made by video
conference calls between Area Chairs and the Program Chairs. For more details
on the reviewing process and the timelines involved see
Appendix~\ref{app:review-details}. 

As Program Chairs of the 2014 conference we decided to test the
consistency of the peer review process through a randomised experiment.
From the 1,678 submissions we chose about 10\% or 170 papers to undergo review by
two separate committees. Each committee was formed by separating the
reviewing body randomly into two groups, while the Area Chairs were
split manually to ensure proper coverage of expertise in the two bodies.
Each selected paper went through the review process
independently. Authors were notified if their paper was in the
experiment as they would have to write two independent author
rebuttals. For the final conference a paper was accepted if any of the
committees would have accepted it. 

\subsection{Outcome Speculations}

\begin{figure}[htb]
\begin{center}
\includegraphics[width=0.40\textwidth]{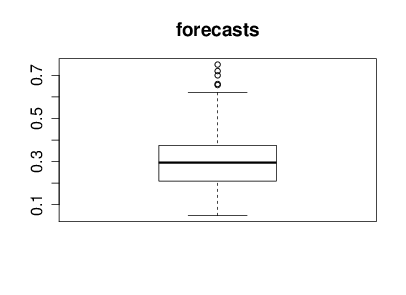}
\end{center}
\caption{Summary forecast from those that responded to a SciCast question about how consistent the decision making was likely to be.}
\label{scicast-forecast}
\end{figure}

Inconsistency in the reviewing process can be quantified in a number
of ways. As the Program Chairs, we opted for asking the question what is the
\% of the papers that yield \emph{inconsistent} decisions.
Before the results were known, with the help of Nicol\'o Fusi, Charles Twardy and
the SciCast team\footnote{SciCast was a collaborative platform for science and technology forecasting created by George Mason University.} we launched a
SciCast question about a week before the results were revealed.  The comment
thread for that question had a lively discussion before the
meeting. Unfortunately, Scicast is currently in hiatus, so we don't have access to that information, but do we have the box plot summary of predictions as shown in Figure~\ref{scicast-forecast}. 

We also made our own predictions.  Corinna forecast this figure would be
25\% and Neil forecast it would be 20\%. 

We see in Figure~\ref{scicast-forecast} that the voting population also perceived that there was likely some inconsistency in the 
reviewing process and the median vote was around 30\%.

\subsection{Results}

The results of this NeurIPS Experiment are summarised in Table
\ref{table-neurips-experiment-results}. Four papers had to
be withdrawn or were rejected without completing the review process
resulting in 166 papers completing the experiment. Of those the two committees disagreed
on 43 or 25.0\% of the papers, which is broadly in line with the speculations above.

\begin{table}[htb]
\caption{Table showing the results from the two committees as a
  confusion matrix. Four papers were rejected or withdrawn without
  review resulting in a final of 166 papers.}
\label{table-neurips-experiment-results}
\centering
\begin{tabular}{lc|c|c|}
& & \multicolumn{2}{c}{Committee 1} \\
& & Accept & Reject \\ \hline
\multirow{2}{*}{Committee 2} & Accept & 22 & 22 \\
& Reject & 21 & 101 
\end{tabular}
\end{table}

Another way of characterizing the outcome is to say that of the papers
accepted, the `other' committee would have only accepted about 50\% of them:
of the 43 accepted papers by Committee 1, Committee 2 only accepted
22, or 51\%, while of the 44 accepted papers by Committee 2, Committee
1 only accepted 22 or 50\% of them. That is, if the conference
reviewing had been run with a different committee, only half of the
papers presented at the conference would have been the same. For other
ways to characterize the result table, see Appendix~\ref{app:neurips-experiment-results}.

\begin{figure}[htb]
\centering
\includegraphics[width=0.45\textwidth]{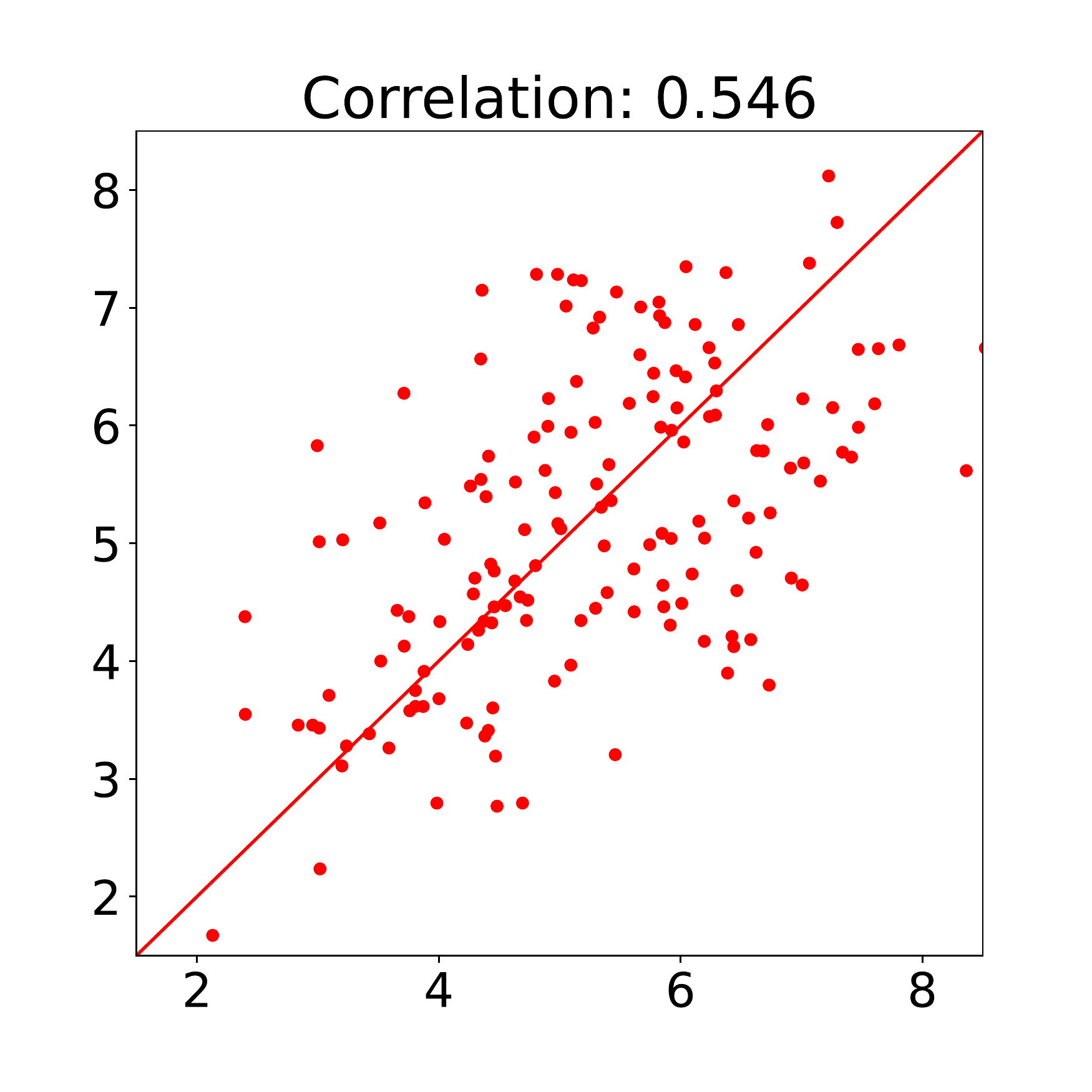}

\caption{Correlation between mean calibrated reviewer scores per paper
  across the two
  independent committees. Standard error on the correlation for
  $n=166$ papers and Gaussian assumptions is $s_r = 0.065$.}
\label{figure-calibrated-quality-correlation}
\end{figure}

In the conference review process, the reviewer verdict is summarized by the `Quantitative Evaluation'
score,\footnote{Somewhat confusingly we will refer to the main score from the `Quantitative Evaluation' as a \emph{quality} score, because it rates the quality of the paper.} see Appendix~\ref{app:review-details}, on a 10 point Likert
scale.  This score is often calibrated per reviewer to account for
difference in reviewer interpretation, see
Section~\ref{sec:calibration}. We looked at the correlation between
the mean calibrated review
scores per paper across the two independent committees. A scatter plot of these
scores from the committees is shown in Figure~\ref{figure-calibrated-quality-correlation}, the Pearson correlation
was computed as $\rho=0.55$. 

During the experiment, the timing of submitted reviews was also
tracked. There is evidence that reviews received after the review submission
deadline, were shorter, gave higher quality scores but with lower
confidence (see Appendix~\ref{app:effect-of-late-reviews}), but there
was insufficient power in the experiment to determine whether this had
a significant effect on the correlation across the program committees.

\section{Public Response}
There was a lot of discussion of the result, both at the
conference and on bulletin boards since. The main topic of discussion
was that the committees were only in agreement on around 50\% of the papers
they accepted. As quickly pointed out by many of the blog posts listed
below, the committees acted better than random, but not by a very large margin. The NeurIPS conference typically has an acceptance rate
of 23.5\% in which case two random committees would be in agreement on
paper decisions only of the order 64\% of the time, or for accepted papers only 
23.5\% of the time. See Appendix~\ref{a-random-committee-25} for a comparison of a random committee with the experiment result.

Public reaction after experiment is documented in a blog post from the time\footnote{\url{https://inverseprobability.com/2015/01/16/blogs-on-the-nips-experiment}} and it seems fair to summarize that some were surprised at the level of inconsistency (despite the pre-experiment speculations falling roughly in line with eventual outcome). The conference itself was run in a very open way, code\footnote{\url{https://github.com/lawrennd/nips2014}} and blog
  posts\footnote{\url{https://inverseprobability.com/2014/12/16/the-nips-experiment}} are all available documenting the decision process.  

Particular public reaction was triggered by a blog post from Eric Price,\footnote{\url{http://blog.mrtz.org/2014/12/15/the-nips-experiment.html}} where much of the discussion speculated on the number of consistent accepts in
the process. 

\section{Analysis}
Having given an overview of the experiment, we now follow up with our
three separate treatments of the results. First we will explore the
relationship between the conference calibration and the experimental
outcome.

\subsection{Reviewer Calibration}
\label{sec:calibration}
NeurIPS papers are evaluated by quality scores on a 10 point Likert
scale (see Appendix~\ref{app:review-details}). A
classical challenge with such scales is that they may be interpreted
differently by different reviewers. Since at least 2005, NeurIPS
chairs have calibrated reviewer scores using scripts of their
own devising. For example, John Platt who chaired the conference in
2006, used a regularised least squares model \citep{Platt-calibration12}. In 2013, Zoubin Ghaharamani
and Max Welling used a Bayesian extension of this model
\citep{Ge-bayesian15}. More recently, outside the NeurIPS community, \cite{MacKay-calibration17} have proposed a Bayesian approach that takes confidence scores into account. 

Like Welling and Ghahramani, we also used a Bayesian variant of the Platt-Burges model, but one that was
formulated as a Gaussian process. We give the details of this approach
in the supplementary material (Appendix~\ref{app:reviewer-calibration}),
but in essence the core of the model is as follows. Each review score
is decomposed into three parts,
$$
y_{i,j} = f_i + b_j + \epsilon_{i, j},
$$
where $y_{i,j}$ is the score from the $j$th reviewer for the $i$th
paper. The score is then decomposed into $f_i$ which is the
\emph{objective} quality of the $i$th paper, i.e.\ it represents the
portion of the score that is common to all the reviewers. The term
$b_j$ is specific to the $j$th reviewer and it represents an offset or
bias associated with the $j$th reviewer. The idea being that different
reviewers interpret the scale differently. Finally $\epsilon_{i,j}$ is
a \emph{subjective} estimate of the quality of paper $i$ according to
reviewer $j$. It reflects how a specific reviewer's opinion differs
from other reviewers. These differences in opinion may be arising due
to differing expertise or perspective.

The model contains $n$ + $m$ + $n\hat{k}$ parameters where $n=1,678$
is the number of papers, $m=1,474$ is the number of reviewers and
$\hat{k}$ is the average number of reviewers per paper. Given that the
data consists of $n\hat{k}$ reviewing scores, the model is
over-parameterised. The original Platt-Burges model used
regularisation to deal with this parameterisation, both
\cite{Ge-bayesian15} and we deal with extra
parameters by allocating them a probability distribution. We use a Gaussian probability resulting in a latent
variable model that has a marginal likelihood which is jointly
Gaussian, so we have
$$
\mathbf{y} \sim N(\mu \mathbf{1}, \mathbf{K}),
$$
where $\mathbf{y}$ is a vector of stacked scores $\mathbf{1}$ is
the vector of ones and the elements of the covariance function are given
by
$$
k(i,j; k,l) = \delta_{i,k} \alpha_f + \delta_{j,l} \alpha_b + \delta_{i, k}\delta_{j,l} \sigma^2,
$$ where $i$ and $j$ are the index of one paper and reviewer and $k$
and $l$ are the index of a potentially different paper and
reviewer. The three parameters of this distribution, $\alpha_f$,
$\alpha_b$, $\sigma^2$ represent the explained variance of the the
score coming from objective quality rating, reviewer offset and
subjective quality rating respectively. As described in the appendix,
the calibrated reviewer score is estimated as the conditional density
of $f_i + \epsilon_{i,j}$. Note that the calibrated reviewer score
\emph{includes} the reviewer's \emph{subjective} opinion about the
paper. See Appendix~\ref{app:reviewer-calibration} for more details
on the model. The parameters of the fitted model are given in
Table \ref{table-fitted-calibration-parameters}.

\begin{table}[htb]
  \label{table-fitted-calibration-parameters}
  \caption{Fitted parameters of the calibration model. The parameters
    are very well-determined as the model is based on around 6,000
    reviewer scores. Once the individual reviewer offset,
    $\alpha_b=0.24$, is removed, the calibrated score $f_i = 1.28$
    plus $\epsilon_{i,j}=1.27$ is made up approximately of subjective
    and objective assessment in roughly equal proportion.}
  \begin{center}
  \begin{tabular}{ccc}
    $\alpha_f$ & $\alpha _b$ & $\sigma^2$ \\ \hline
    1.28 & 0.24 & 1.27
  \end{tabular}
  \end{center}
\end{table}  

Under the model assumptions we see that calibrated review scores are
made up of subjective and objective opinion in roughly equal
proportions. In other words, 50\% of a typical reviewer's score is
coming from opinion that is particular to that reviewer and \emph{not}
shared with the other reviewers. This figure may seem large, but in
retrospect it is perhaps not surprising. Papers are judged by
subjective criteria such as novelty as well as more objective criteria
such as rigour. The subjectivity of reviewer scores also seems a
sensible starting point to  uncover the inconsistency between the two
committees described by the NeurIPS experiment. 

The result is consistent with the correlation coefficient we computed
between the two independent committees, $\rho = 0.55 \pm 0.065$. Our
calibration model is suggesting that the overall correlation between
two committees would be given by $0.502 = 1.28/(1.28+1.27)$.

\section{Simulation of Subjective Scoring}\label{sec:simulation-of-subjective-scoring}

\begin{figure}[htb]
\centering
\includegraphics[width=0.65\textwidth]{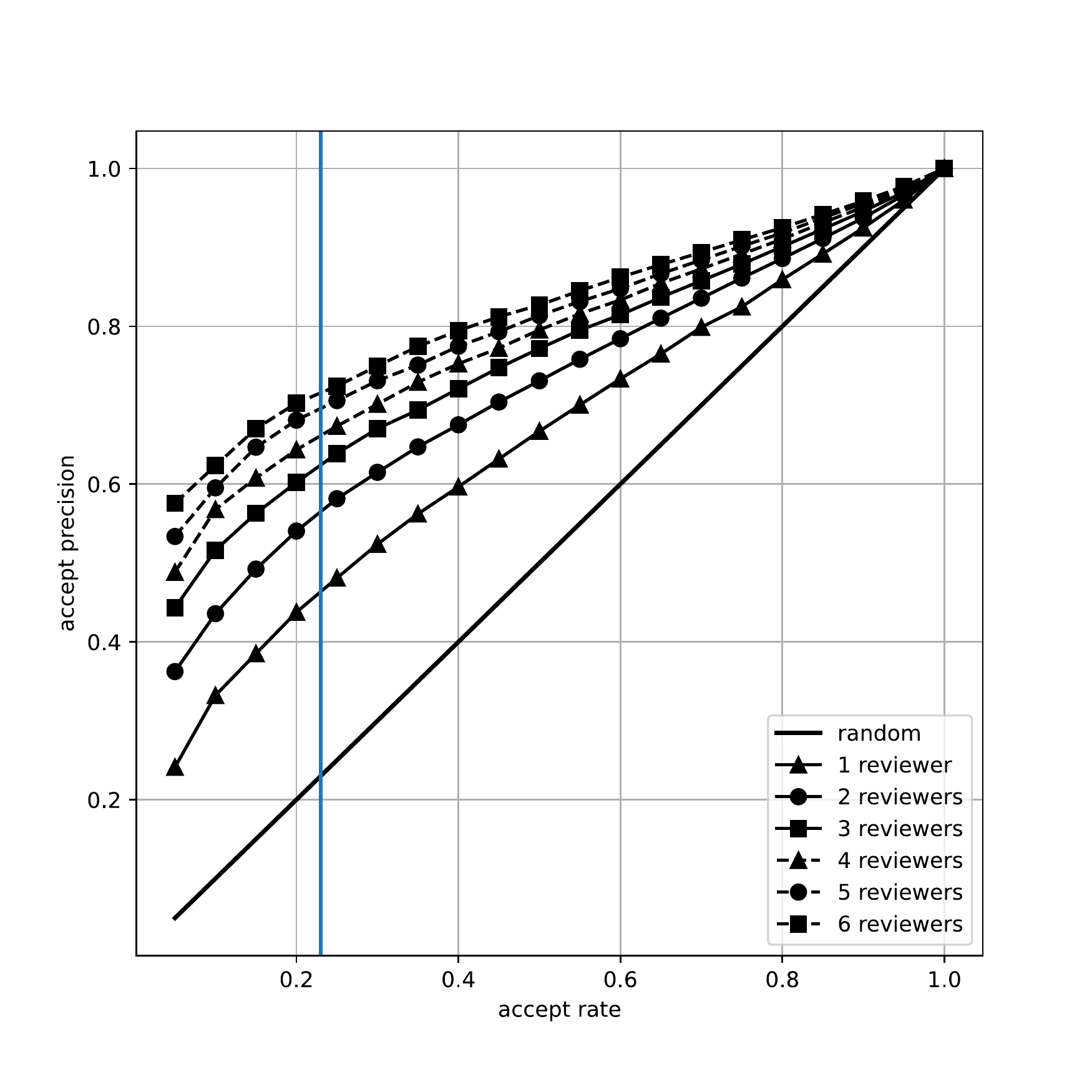}

\caption{Plot of the accept rate versus the accept consistency of the
  conference for 50\% subjectivity in a simulation of the conference
  with different numbers of reviewers per paper. We've marked a vertical line at the NeurIPS 2014 accept rate of 23\%.}
\label{figure-consistency-vs-accept-rate}
\end{figure}

To check whether this subjective scoring also explains the
inconsistency in acceptance decisions between the two committees, we set up a
simple simulation study.\footnote{The code for running the simulation can be found at \url{https://github.com/lawrennd/neurips2014/blob/master/notebooks/neurips-simulation.ipynb}} For our simulations, we assumed that each
paper was scored according to the model we've given above and we
estimated the accept consistency through averaging across 100,000
samples. In Figure~\ref{figure-consistency-vs-accept-rate} we show the
estimates of the accept consistency as a function of conference accept
rate. For three reviewers and 50\% subjectivity, the simulation
suggests that we should expect an accept consistency of around
63\%. This is higher than the accept consistency that we observed, but
it falls within the bounds of statistical error suggested e.g.\ by a
Bayesian analysis of the data (see Appendix~\ref{bayesian-analysis}). Conceptually, the large error comes because
although experimental sample size overall is a relatively healthy count of 166,
the low accept rate for the conference (in 2014 it was 23\%) means that the number of
samples when exploring consistency across the two committees is around
40. This leads to a standard error of our
estimate\footnote{Consider Committee 1 with $n=43$ accepts. Of these
  21 are rejected by committee 2. Our estimate of the probability of
  inconsistency is given by $p=21/43$ with a standard error of
  $\sqrt{\frac{p(1-p)}{n}} = 0.076$.} of around 8\%.  Of course, the
simulation model also oversimplifies the complexity of the final
decision process, which involved detailed discussion of each papers
between reviewers authors and program committee members. This
simplification may be introducing some optimistic bias in the
simulation's estimate of the final accept precision. But it seems that
our model simulation model is a plausible starting point for exploring
the expected consistency of a given reviewing set up.

The simple simulation we describe suggests that a major source of
inconsistency in the conference can be traced back to subjectivity in
the reviews. Combination of the calibration model and our simulation
suggests an accept precision for the conference of around 61\%. This
is consistent with the upper end of consistency estimates: analysis in Appendix~\ref{uncertainty-accept-rate}
suggests that the accept precision was was between 38\% and 64\%. This highlights the
unreliability of the accept precision statistic. The statistical power
is low because the number of samples used in its calculation is given
by accept rate $\times$ experiment sample size.

\subsection{Consistency and Correctness}

It seems self-evident that we are looking for greater consistency between review
committees. After all, if decisions are inconsistent, then how can
they be `correct'? While it's true that inconsistency implies
incorrectness, the converse is not true. Consistency does not imply
correctness. For example, if both committees were to choose papers to
accept based on how many references they include, then their decisions
would be consistent, but not correct. Given that we know that
\emph{incorrect} decisions will be made, then we can also phrase the
question in another way. Given that there will be errors, do we prefer
errors which will be consistently made? When there are errors,
variation in decision making may be a good thing: it could prevent a
particular type of paper being consistently discriminated against.

We've established that there is inconsistency in the peer review
process, and we have associated that inconsistency with subjective
scoring by reviewers. But we have also warned against over emphasis on
consistency as an aim for a reviewing process. Consistency is only a
good thing if the decisions can also be shown to be correct. So, a
follow up question would seem to be: how good is the committee at
selecting the `right' papers? Unfortunately we don't have a ground
truth assessment of what the right papers are. But, because time has
passed between the conference and today, we can explore what happened
to accepted papers in terms of their citation impact.

\section{Impact of Accepted Papers}

Seven years have passed since the NeurIPS experiment and the papers
published at the conference have had time to establish themselves. In
this section we explore how they fared in terms of their
\emph{citation impact}.\footnote{Citation counts certainly have flaws
  when being used as a measure of impact of a paper. They do not
  represent, e.g. adoption in industry, adoption for public policy and
  public awareness/education. However, they do provide a quantitative
  measure that allows us to analyse the impact of conference papers at
  scale.}

To determine the citation impact of papers, we searched for all
accepted papers (not just the once in the experiment) from the conference on Semantic Scholar.\footnote{See
  \url{https://www.semanticscholar.org/about}.}
The Semantic Scholar ID of the papers was recorded and we made use of
the Semantic Scholar API to retrieve the number of citing papers. The
citation scores were transformed into the citation impact using a
monotonic transformation as follows,
$$
\text{citation impact} = \log_{10} (1 + \text{number of citations}).
$$
This transformation eliminated the heavy tails of the distribution
of citations, leading to a citation score distribution that is closer
to a Gaussian, enabling us to make use of Pearson's $\rho$ for
correlation measurement.

\begin{figure}[htb]
  \begin{center}
    \includegraphics[width=0.5\textwidth]{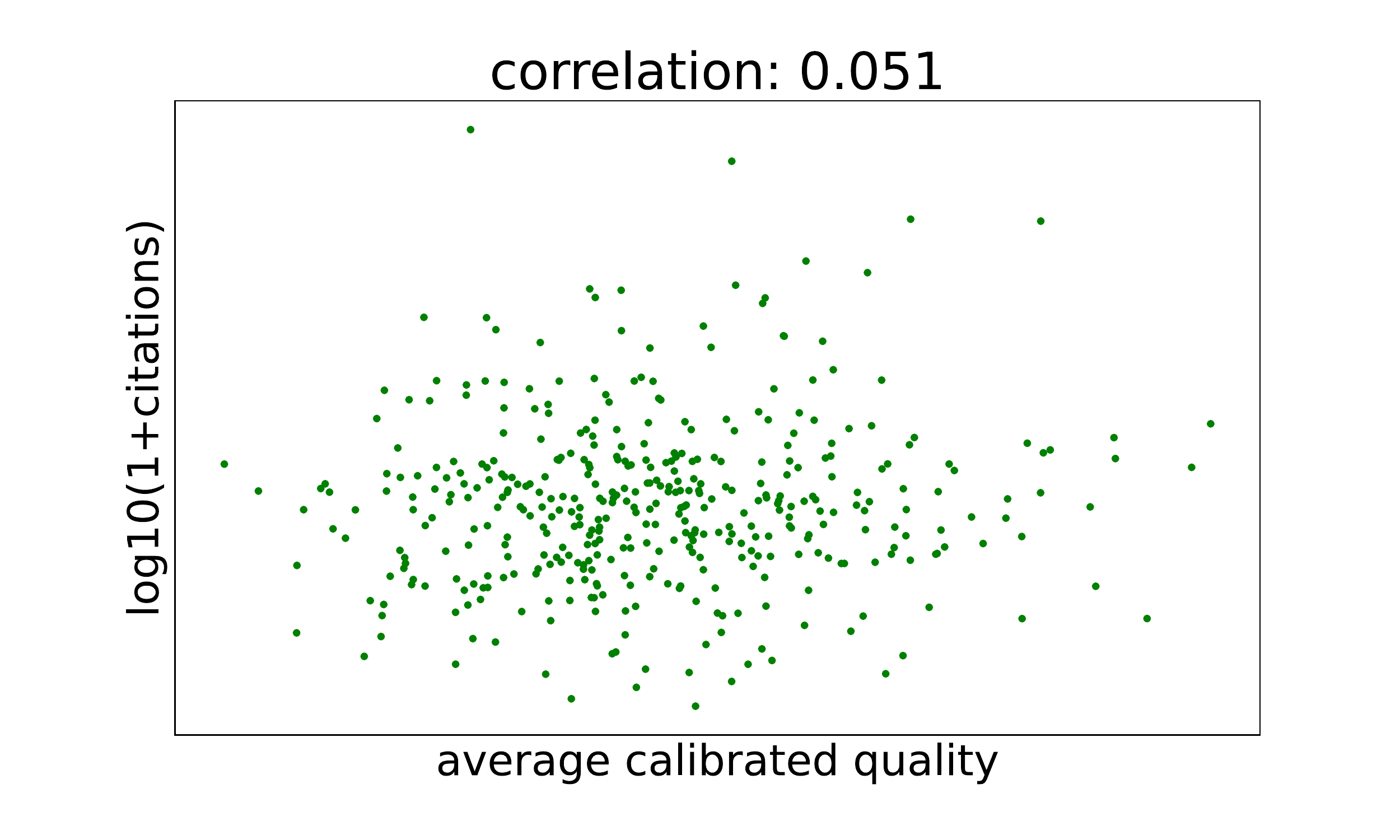}
  \end{center}
  \caption{Scatter plot of the citation impact (defined as
    $\log_{10}(1+\text{citations})$) against the average calibrated
    quality score for accepted NeurIPS 2014 papers. To prevent
    reidentification of an individual paper's quality scores each
    point is corrupted by differentially private noise in the plot
    (correlation is computed before adding differentially private
    noise). We have also purposely left off the scale, as the main
    point in including the scatter plot is to show the general shape
    of the points, validating our use of Pearson's correlation
    coefficient, $\rho$. The sample size is 414 accepted papers which
    gives us $\rho = 0.051 \pm 0.049$ indicating no significant
    correlation.}
  \label{figure-citations-vs-average-calibrated-quality-accept}
\end{figure}

We computed the correlation between the average calibrated quality
score and the citation impact. There was \emph{no significant
  correlation} between those scores. We show a scatter plot of the
data in Figure~\ref{figure-citations-vs-average-calibrated-quality-accept}. In the
scatter plot we have added differential privacy noise to the values
 to obfuscate individual paper
identities. Correlation coefficient is computed before adding the
differential privacy noise (see Appendix~\ref{correlation-of-quality-scores-and-citation}).

The calibrated quality score is not specifically designed to measure
impact, but it still may be surprising that there's no correlation
between this score and citation impact for the group of accepted
papers. The implication that the quality score, which is the main
criterion on which accept/reject decisions are being made, is
uninformative in determining the paper's eventual influence should
give pause for thought.

Does this mean that reviewers can't judge what papers are likely to be
influential? Or is something else going on? In 2013 Welling and
Ghahramani introduced a separate scoring indicator. 
Perhaps the answer to our quandary in the quality
scores lies in the phrasing of the question they introduced ((see also
Appendix~\ref{app:review-details}).
\begin{quote}
  Independently of the Quality Score above, this is your opportunity to
identify papers that are very different, original, or otherwise
potentially impactful for the NIPS community.
\end{quote}

\begin{figure}[htb]
  \begin{center}
    \includegraphics[width=0.5\textwidth]{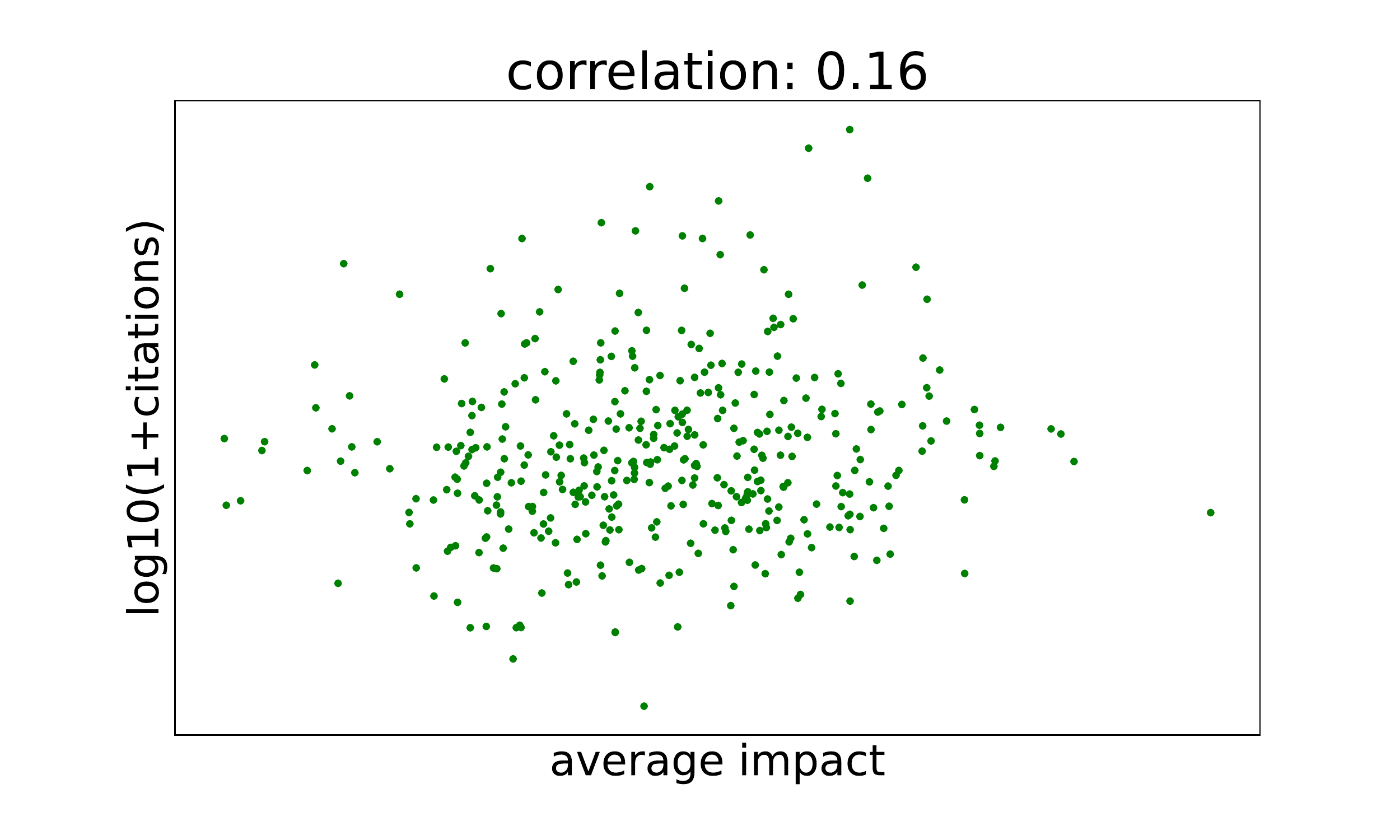}
  \end{center}
  \caption{Scatter plot of the citation impact (defined as
    $\log_{10}(1+\text{citations})$) against the average impact score
    for accepted NeurIPS 2014 papers. As in Figure~\ref{figure-citations-vs-average-calibrated-quality-accept}, data
    is corrupted for plotting by differentially private noise. The
    scatter plot is to show the general shape of the points,
    validating our use of Pearson's correlation coefficient,
    $\rho$. Correlation is $0.16 \pm 0.049$ giving a statistically
    significant result.}
  \label{figure-citations-vs-average-impact-accept}
\end{figure}

Here reviewers are being asked to judge the likely future impact of
the paper. The score is a binary value, work is categorised as being
either `potentially have a major impact' or `unlikely to have much
impact'. Analysis of this score does show a statistically significant
correlation with an accepted paper's citation impact (see Figure~\ref{figure-citations-vs-average-impact-accept}), but the magnitude of
the effect is small.

Also note that the correlation for this score across the duplicated
papers in the NeurIPS experiment was relatively low:\footnote{Standard
  error compute as $\sqrt{\frac{1-\rho^2}{n-2}}$.} $0.27 \pm 0.075$.

Alongside quality and impact, reviewers are asked to provide a
confidence score for their review. Scored on a Likert scale between 1
(`the review is an educated guess') and 5 (`the reviewer is absolutely
certain'). The confidence score helps Area Chairs decide how much
weight to place on a particular review and whether new reviews may be
needed for a particular paper. Appendix~\ref{confidence-score} gives a
detailed description of the scale.

The questions about reviewer confidence are entirely focused on how a
given reviewer feels about a particular paper, so they reflect an
individual reviewers expertise. But, interestingly, the confidence
score is the most predictive of the paper's final impact. This implies
that underlying the confidence score there is also a particular
property of the paper being represented. Likely, reviewers are more
confident about well written papers that clearly express the core
ideas in the paper. The confidence score may be reflecting some
underlying clarity of the paper. Such clarity is also likely to have a
downstream effect on citation impact.

\begin{figure}[htb]
\centering
\includegraphics[width=0.5\textwidth]{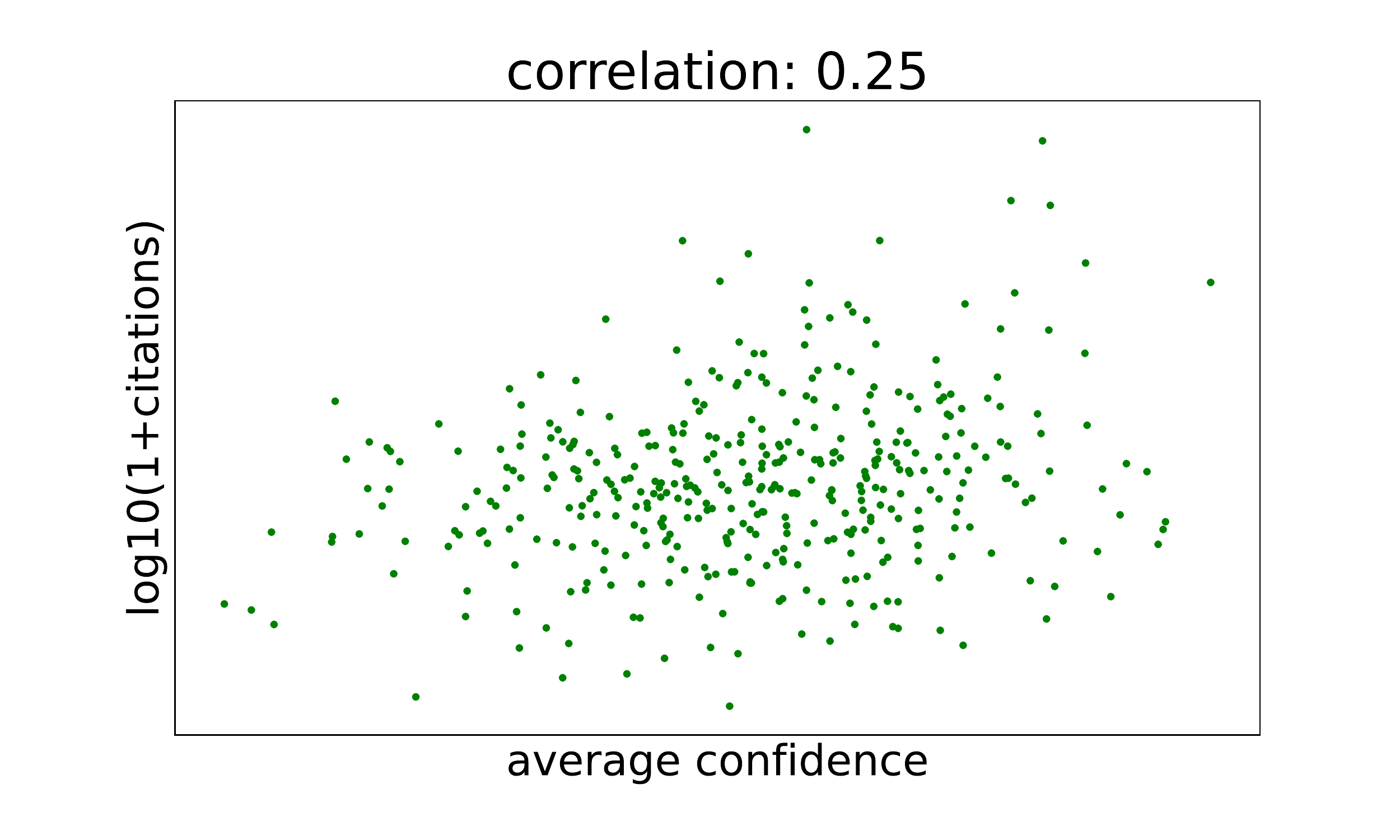}

\caption{Scatter plot of $\log_{10}(1+\text{citations})$ against the
  average confidence score for accepted papers. As in Figure~\ref{figure-citations-vs-average-calibrated-quality-accept}, data is
  corrupted for plotting by differentially private noise. The scatter
  plot is to show the general shape of the points, validating our use
  of Pearson's correlation coefficient, $\rho$. Correlation is $0.25
  \pm 0.048$ giving a statistically positive correlation between the
  two values.}
\label{figure-citations-vs-average-confidence-accept}
\end{figure}

More evidence for this notion of clarity comes from analysing the
correlation of the confidence score across the two different
committees in the NeurIPS experiment. Correlation of confidence scores
was $0.39 \pm 0.072$, giving additional evidence for
the influence of paper clarity in the reviewers' confidence about the paper.

Conferences and journals are often measured and ranked by their impact
factor. These impact factors are metrics that derive from citation
counts of the papers presented in those outlets. But our long term
analysis shows that at NeurIPS 2014, quality scores were uncorrelated
with the final citation impact of the papers. This raises serious
questions about our processes. For accepted papers, the reviewers' notion of quality is independent of the papers' final
influence on the field. Indeed, reviewer instructions for the
conference implied that should be the case, with a request for an
additional impact score that is independent of the quality score.

We were motivated to explore the relation between quality scores and
citation counts in an effort to determine how `correct' the decisions
were within the reviewing process. We have argued that if errors are
being made, then we would be better off making such errors
inconsistently, rather than always rejecting the papers for the same
misconceptions about the role of the reviewing process. If we accept
that final paper citation counts are some measure of paper quality,
then we see that reviewers fail to capture this in their scores.

Finally, reviewer confidence scores are being influenced by particular
characteristics of the papers, what we might think of as the clarity
of the paper. As a result, in our analysis, reviewer confidence turned
out to be the best indicator of the paper's citation impact.

We'll return to the implications of this analysis in the discussion,
but before exploring further, we will turn our attention to the
\emph{rejected} papers. Accepted papers, by their nature, are those
that are scoring at the high end of the quality score. To explore the
low end of the quality score, we turn to papers that were not
presented at the conference.

\section{Fate of Rejected Papers}

Of the 1,678 papers submitted to NeurIPS 2014, only 414 were presented
at the final conference. In the previous section we have shown that
the reviewer-assessed quality of these papers gave no indication of
their final impact on the community as measured by citation
counts. But what about the rejected papers?

\begin{figure}[htb]
\centering
\includegraphics[width=0.5\textwidth]{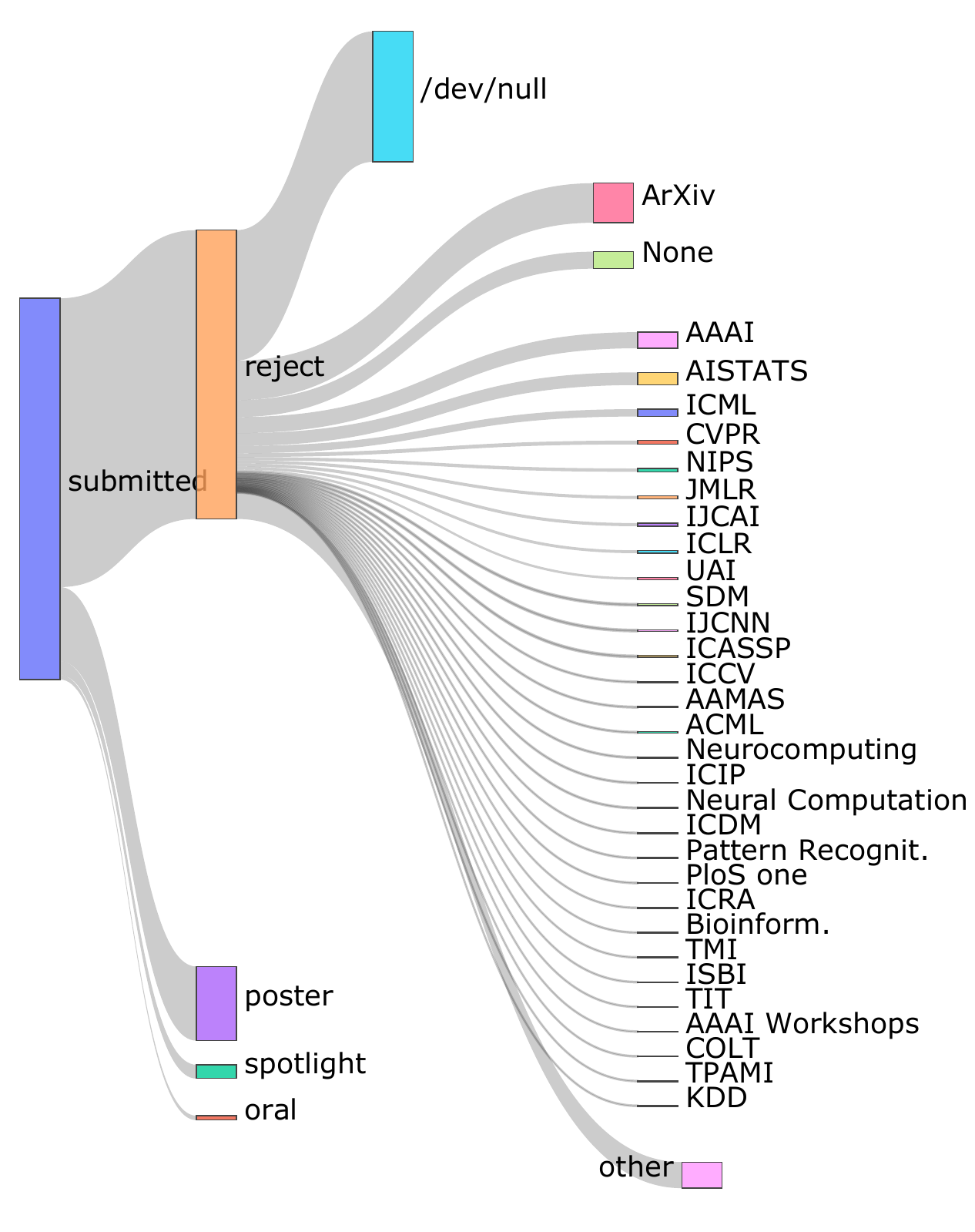}

\caption{Sankey diagram showing the flow of NeurIPS papers through the system from submission to eventual publication.}
\label{figure-where-do-neurips-papers-go}
\end{figure}

To trace the fate of the rejected papers, we searched Semantic Scholar
for evidence of all 1,264 rejected papers. We looked for papers with
similar titles and where the NeurIPS submission's contact author was
also in the author list. We were able to track down 680 papers. Of
these 177 were only found on arXiv, 76 were found as PDFs online
without a publication venue and 427 were published in other
venues. The outlets that received ten or more papers from this group
were AAAI (72 papers), AISTATS (57 papers), ICML (33 papers), CVPR (17
papers), Later NeurIPS (15 papers), JMLR (14 papers), IJCAI (14
papers), ICLR (13 papers), UAI (11 papers).  Opinion about quality of
these different outlets will vary from individual, but from our
perspective all of these outlets are `top-tier' for machine learning
and related areas. Other papers appeared at less prestigious outlets, and citation scores were also recored for papers that remained available only on ArXiv.  Note that there is likely a bias towards outlets
that have a submission deadline shortly after NeurIPS decisions are
public, e.g.\ submission deadline for AAAI 2015 was six days after
NeurIPS decisions were sent to authors. AISTATS has a submission
deadline one month after. A Sankey diagram showing where papers
submitted to the conference ended up is shown in Figure~\ref{figure-where-do-neurips-papers-go}.

\begin{figure}[htb]
\begin{center}
\includegraphics[width=0.5\textwidth]{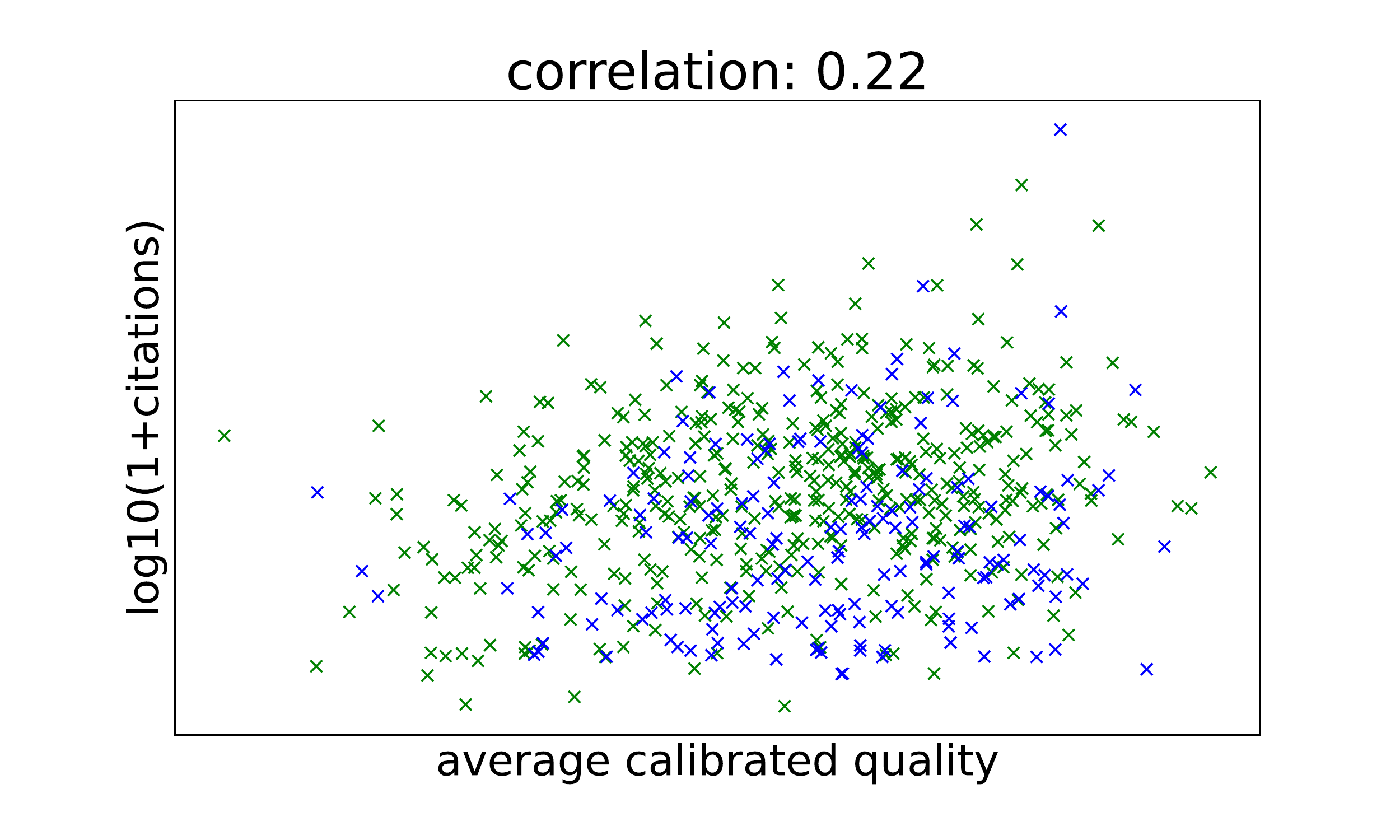}
\end{center}
\caption{Scatter plot of $\log_{10}(1+\text{citations})$ against the
  average calibrated quality score for rejected papers. To prevent
  reidentification of individual papers quality scores and citation
  count, each point is corrupted by differentially private noise in
  the plot (correlation is computed before adding differentially
  private noise). Correlation is $0.22 \pm 0.038$. }
\label{figure-citations-vs-average-calibrated-quality-reject}
\end{figure}

For each paper we found we used the Semantic Scholar ID to recover the
citation count for the paper. This allowed us to measure the
correlation between the rejected papers' quality scores and their
eventual citation impact. The scatter plot is shown in Figure~\ref{figure-citations-vs-average-calibrated-quality-reject}.

The results show weak correlation between the rejected paper's quality
scores and their citation impact. It seems that for rejected papers the reviewing body's quality scores do have some correlation with citation impact. 

In Appendix~\ref{correlation-of-quality-scores-and-citation} we plot rejected papers and accepted papers together (Figure \ref{figure-citations-vs-average-calibrated-quality-all}). Among the rejected papers are papers which received hundreds and thousands of citations. The most highly cited rejected paper has more citations than all but two of the accepted papers. 

\section{Conclusions}

In this paper we have revisited the 2014 NeurIPS Experiment. The
experiment investigated the consistency of peer review by randomly
selecting 10\% of the papers submitted to the conference and having
them reviewed by two separate committees. The experiment found that
(roughly) 25\% of decisions were inconsistent between the two
committees. Reaction to the experiment in the community focused on a
statistic we call the accept precision. This statistic represents the
percentage of presented papers that would have been the same through
an independently rerun version of the same reviewing process. Our analysis has shown this statistic to be
unreliable, where a headline figure of 50\% has been presented, the  Bayesian analysis in Appendix~\ref{bayesian-analysis} suggests that the actual figure is somewhere between 38\% and 62\%.

By revisiting the conference calibration process, we explored the
effect of subjective reviewer opinion on the accept consistency. The
calibration process suggested that around 50\% of the variance in
reviewer scores is associated with subjective opinion. We built a
simple simulation that used this figure in reviewer scoring, this
simulation suggested that when each paper has three reviewers, if the
underlying subjectivity is 50\%, we expect an accept precision of
62\%. The simulation showed that we can improve this consistency with
more reviewers, but there are diminishing returns as the number of
reviewers increase.

Having reviewed the consistency of the conference, we then emphasised
that consistency does not mean correctness. It is easy to have a
conference where the decisions between two independent committees
would be consistent but arbitrary. For example, a conference where papers
were accepted on the basis of the number of references they made. We
therefore explored the extent to which the reviewer quality scores are
predictive of eventual paper impact.

By determining the number of citations for each accepted paper, we
were able to demonstrate that there is no correlation between reviewer
quality scores for accepted papers and the papers' eventual citation
impact. Given that papers were accepted on the basis of their quality
scores, this raises questions as to what the objective of the
reviewing process is. Welling and Ghahramani had introduced a new
impact score for the 2013 conference that we
continued to use. This score did have some correlation with the
eventual citation impact although the largest correlation arose from
examining reviewer confidence. We speculated that this may be to do
with an underlying element of clarity in the papers that gives
reviewers confidence and makes papers more likely to be cited.

Finally, we followed up the fate of papers that were rejected from the
conference. We found that a very significant portion of rejected
papers are published very soon after the NeurIPS conference in other
high quality venues. We found that there was weak correlation between rejected papers' citation impact and their quality scores, suggesting that
reviewers are better at judging which papers are weak, in terms of
citation impact, than they are at judging which papers are strong.

In summary, we suggest a significant overhaul of the scoring process
of machine learning conferences. It is clear that reviewers are
picking up on at least three different components for a paper. There
is a notion of `quality', that in its upper range seems to be
independent of citation impact. There is a notion of `impact' that is
only weakly correlated with the measured citation impact, and we
speculate that there is also a notion of clarity: reviewers were more
confident about papers that later turned out to have a high citation
impact.

The notion of `quality' for the NeurIPS conference may be conflating
separate ideas. In particular, reviewer instructions indicate that
rigour is also a component of the quality score. If this is the case
then it may explain some of the inconsistency between
reviewers. We also note the inconsistency between the instructions `Qualitative Evaluation' description (see Appendix~\ref{qualitative-evaluation}), which is nicely partitioned into separate descriptions for `Quality', `Clarity', `Originality' and `Significance' and the `Quantitative Evaluation' (see Appendix~\ref{quantitative-evaluation}) which summarises the paper with a single score. Perhaps some of the subjective opinion in scoring is arising
from the different weightings different reviewers may be placing on
these different underlying factors. For future conferences, we suggest that scores
could be separated out to improve consistency between reviewers. For
example reviewers could be asked to score across clarity, rigour,
significance and originality. Final rankings could then be determined
through some combination of these scores that would be agreed at
program committee level, rather than at the whim of individual
reviewers.

As the size of the community has grown, and we are all becoming less familiar with each others individual research contributions, there is a danger that the number of publications at 'top-tier' conferences becoming a proxy to represent the quality of an individual researcher. For early career researchers, who've had less time to establish themselves, this proxy measure will be highly sensitive to inconsistency in reviewing processes. With increasing commercial interest in machine learning, 'top-tier' conference publication has also begun to creep into corporate performance review processes. Again, if the performance review is taking in a shorter period of time, this measure will be highly sensitive to inconsistency in the review process. Given that we've shown that such inconsistencies exist, we would suggest that we should be vary wary of 'top-tier' publication counts as a measure of individual researcher quality.

We understand that the 2021 NeurIPS Program Chairs are repeating the experiment and look forward to seeing how their results compare to ours.

\subsubsection*{Acknowledgements}

We would like to thank the authors, reviewers, Area Chairs of NeurIPS 2014 for not just carrying the writing and reviewing burden, but also the additional work load of reviewing some of the papers twice. We also thank Max Welling and Zoubin Ghahramani, the 2014 General Chairs, as well as the NeurIPS Board for supporting us in the experiment. Finally, we thank the maintainers of CMT for accommodating the experiment with the system and the Semantic Scholar team at AI2 for making available an API for the important work of exploring the impact of academic publishing.

\bibliographystyle{plainnat}
\bibliography{revisiting-the-neurips-experiment}
\appendix
\section{Details of the Reviewing Process}
\label{app:review-details}

In this section, for reference, we provide some details on the reviewing process for NeurIPS 2014. We include time line, and the paper scoring instructions. Reviewer instructions and paper scoring evolve over the years for NeurIPS. Broadly speaking, the instructions for the 2014 edition of the conference remained the same as the 2013 edition, with additional stipulations given for those who's papers were selected for the NeurIPS experiment.

The timeline for the conference is given in Table \ref{table-reviewing-timeline}.

\begin{table}[hb]
    \centering
    \caption{Timeline for the 2014 conference. Dates are each given in ISO format. Timeline broadly matched that used in previous years of the conference.}
    \label{table-reviewing-timeline}
    
    \begin{tabular}{l|r}
    Event & Date \\ \hline
    
        Submission Deadline & 2014-06-06  \\
        Bidding Open for Area Chairs (this was \emph{delayed} by
  CMT issues) & 2014-06-12 \\
  Bidding Open for Reviewers & 2014-06-17 \\
  Start Reviewing & 2014-07-01 \\
  Reviewing deadline & 2014-07-21 \\
  Reviews to Authors & 2014-08-04 \\
  Author Rebuttal Due & 2014-08-11 \\
  Teleconferences Begin & 2014-08-25 \\
  Teleconferences End & 2014-08-30 \\
  Preliminary Decisions Made & 2014-09-01 \\
  Decisions Sent to Authors & 2014-09-09
    \end{tabular}
    
\end{table}

\subsection{Paper Scoring and Reviewer
Instructions}\label{paper-scoring-and-reviewer-instructions}

To keep quality of reviews high, we tried to keep reviewing load low. We didn't
assign any reviewer more than 5 papers, most reviewers received 4
papers.
The instructions to reviewers for the 2014 conference are 
available online\footnote{See  
\url{https://neurips.cc/Conferences/2014/PaperInformation/ReviewerInstructions}} and we summarise also below.

The reviewers assign a number of scores: Quantitative Evaluation,
Impact Score, Confidence Score, and a verbal Qualitative Evaluation to
each paper they review. For detail about these 4 assessments, see below.

\subsection*{Quantitative Evaluation}\label{quantitative-evaluation}
Reviewers give a paper score of between 1 and 10 for each paper. The program
committee will interpret the numerical score in the following way:

\begin{description}
\item[10] Top 5\% of accepted NeurIPS papers, a seminal paper for the ages.

  I will consider not reviewing for NeurIPS again if this is rejected.
\item[9] Top 15\% of accepted NeurIPS papers, an excellent paper, a strong
  accept.

  I will fight for acceptance.
\item[8] Top 50\% of accepted NeurIPS papers, a very good paper, a clear
  accept.

  I vote and argue for acceptance.
\item[7] Good paper, accept.

  I vote for acceptance, although would not be upset if it were
  rejected.
\item[6] Marginally above the acceptance threshold.

  I tend to vote for accepting it, but leaving it out of the program
  would be no great loss.
\item[5] Marginally below the acceptance threshold.

  I tend to vote for rejecting it, but having it in the program would
  not be that bad.
\item[4] An OK paper, but not good enough. A rejection.

  I vote for rejecting it, although would not be upset if it were
  accepted.
\item[3] A clear rejection.

  I vote and argue for rejection.
\item[2] A strong rejection. I'm surprised it was submitted to this
  conference.

  I will fight for rejection.
\item[1] Trivial or wrong or known. I'm surprised anybody wrote such a
  paper.

  I will consider not reviewing for NeurIPS again if this is accepted.
\end{description}

Reviewers should NOT assume that they have received an unbiased sample
of papers, nor should they adjust their scores to achieve an artificial
balance of high and low scores. Scores should reflect absolute judgments
of the contributions made by each paper.

\subsection*{Impact Score}\label{impact-score}

The impact score was an innovation introduce in 2013 by Ghahramani and
Welling that we retained for 2014. 

Independently of the Quality Score above, this is your opportunity to
identify papers that are very different, original, or otherwise
potentially impactful for the NeurIPS community.

There are two choices:

\begin{description}
\item[2] This work is different enough from typical submissions to potentially
have a major impact on a subset of the NeurIPS community.

\item[1] This work is incremental and unlikely to have much impact even though
it may be technically correct and well executed.
\end{description}

Examples of situations where the impact and quality scores may point in
opposite directions include papers which are technically strong but
unlikely to generate much follow-up research, or papers that have some
flaw (e.g.~not enough evaluation, not citing the right literature) but
could lead to new directions of research.

\subsection*{Confidence Score}\label{confidence-score}

Reviewers also give a confidence score between 1 and 5 for each paper.
The program committee will interpret the numerical score in the
following way:

\begin{description}
\item[5] The reviewer is absolutely certain that the evaluation is correct and
very familiar with the relevant literature.

\item[4] The reviewer is confident but not absolutely certain that the
evaluation is correct. It is unlikely but conceivable that the reviewer
did not understand certain parts of the paper, or that the reviewer was
unfamiliar with a piece of relevant literature.

\item[3] The reviewer is fairly confident that the evaluation is correct. It
is possible that the reviewer did not understand certain parts of the
paper, or that the reviewer was unfamiliar with a piece of relevant
literature. Mathematics and other details were not carefully checked.

\item[2] The reviewer is willing to defend the evaluation, but it is quite
likely that the reviewer did not understand central parts of the paper.

\item[1] The reviewer's evaluation is an educated guess. Either the paper is
not in the reviewer's area, or it was extremely difficult to understand.
\end{description}

\subsection*{Qualitative Evaluation}\label{qualitative-evaluation}

All NeurIPS papers should be good scientific papers, regardless of their
specific area. We judge whether a paper is good using four criteria; a
reviewer should comment on all of these, if possible:

\begin{itemize}
\item
  Quality

  Is the paper technically sound? Are claims well-supported by
  theoretical analysis or experimental results? Is this a complete piece
  of work, or merely a position paper? Are the authors careful (and
  honest) about evaluating both the strengths and weaknesses of the
  work?
\item
  Clarity

  Is the paper clearly written? Is it well-organized? (If not, feel free
  to make suggestions to improve the manuscript.) Does it adequately
  inform the reader? (A superbly written paper provides enough
  information for the expert reader to reproduce its results.)
\item
  Originality

  Are the problems or approaches new? Is this a novel combination of
  familiar techniques? Is it clear how this work differs from previous
  contributions? Is related work adequately referenced? We recommend
  that you check the proceedings of recent NeurIPS conferences to make sure
  that each paper is significantly different from papers in previous
  proceedings. Abstracts and links to many of the previous NeurIPS papers
  are available from http://books.NeurIPS.cc
\item
  Significance
  
  Are the results important? Are other people (practitioners or
  researchers) likely to use these ideas or build on them? Does the paper
  address a difficult problem in a better way than previous research? Does
  it advance the state of the art in a demonstrable way? Does it provide
  unique data, unique conclusions on existing data, or a unique
  theoretical or pragmatic approach?
\end{itemize}

\section{Code and Data}

All code for our experiments can be found on line in Jupyter notebooks, \url{https://github.com/lawrennd/neurips2014/} and an accompanying python package for dealing with the data \url{https://github.com/lawrennd/cmtutils/}, it can be installed with `pip install cmtutils`. Code was all originally written in Python 2, but is now updated for compatibility with Python 3.

The repository has all the code used for managing the NeurIPS reviewing process in a set of different notebooks. The relevant notebooks for recreating the different experiments below are listed in each section.

Unfortunately, the full data for recreating our analysis is not public, due to challenges around anonymizing reviewer identity and the identity of authors whose papers weren't accepted at the conference.

\section{NeurIPS Experiment Results}
\label{app:neurips-experiment-results}

The results of the experiment are listed in Table~\ref{table-neurips-experiment-results} in the main body of the paper. A Jupyter notebook for recreating this analysis can be found here \url{https://github.com/lawrennd/neurips2014/blob/master/notebooks/The%20NIPS%20Experiment.ipynb}.

There are a few ways of summarising the numbers in this table as percent
or probabilities. First, the inconsistency is the proportion of decisions
that were not the same across the two committees. The decisions were
inconsistent for 43 out of 166 papers or 0.259 as a proportion. This
number is a natural way of summarising the figures if you are
submitting your paper and wish to know an estimate of what the
probability is that your paper would have different decisions according
to the different committees. 

Secondly, the accept precision: if you are
attending the conference and looking at any given paper, then you might
want to know the probability that the paper would have been rejected in
an independent rerunning of the conference. We can estimate this for
Committee 1's conference as 22/(22 + 22) = 0.5 (50\%) and for Committee
2's conference as 21/(22+21) = 0.49 (49\%). Averaging the two estimates
gives us 49.5\%. 

Thirdly, the reject precision: if your paper was
rejected from the conference, you might like an estimate of the
probability that the same paper would be rejected again if the review
process had been independently rerun. That estimate is 101/(22+101) =
0.82 (82\%) for Committee 1 and 101/(21+101)=0.83 (83\%) for Committee
2, or on average 82.5\%. 

A final quality estimate might be the ratio of
consistent accepts to consistent rejects, or the agreed accept rate,
22/123 = 0.18 (18\%).

\begin{itemize}
\tightlist
\item
  \emph{inconsistency}: 43/166 = \textbf{0.259}

  \begin{itemize}
  \tightlist
  \item
    proportion of decisions that were not the same
  \end{itemize}
\item
  \emph{accept precision} \(0.5 \times 22/44\) + \(0.5 \times 21/43\) =
  \textbf{0.495}

  \begin{itemize}
  \tightlist
  \item
    probability any accepted paper would be rejected in a rerunning
  \end{itemize}
\item
  \emph{reject precision} = \(0.5\times 101/(22+101)\) +
  \(0.5\times 101/(21 + 101)\) = \textbf{0.175}

  \begin{itemize}
  \tightlist
  \item
    probability any rejected paper would be rejected in a rerunning
  \end{itemize}
\item
  \emph{agreed accept rate} = 22/101 = \textbf{0.218}
  \begin{itemize}
    \item
  ratio between agreed accepted papers and agreed rejected papers.
  \end{itemize}
\end{itemize}

\subsection{Reaction After Experiment}\label{reaction-after-experiment}

There was a lot of discussion of the result, both at the
conference and on blogs and bulletin boards.\footnote{A summary of posts can be found here \url{https://inverseprobability.com/2015/01/16/blogs-on-the-nips-experiment}} Such discussion is to be
encouraged, and for ease of memory, it is worth pointing out that the
approximate proportions of papers in each category can be nicely divided
in to eighths as follows. Accept-Accept 1 in 8 papers, Accept-Reject 3
in 8 papers, Reject-Reject, 5 in 8 papers. This makes the statistics
we've computed above: inconsistency 1 in 4 (25\%) accept precision 1 in
2 (50\%) reject precision 5 in 6 (83\%) and agreed accept rate of 1 in 6
(20\%). This compares with the accept rate of 1 in 4.

\begin{itemize}
\item
  Public reaction after experiment is documented in a blog post.\footnote{\url{https://inverseprobability.com/2015/01/16/blogs-on-the-nips-experiment}}

\item
  NeurIPS was run in a very open way:
  code\footnote{\url{https://github.com/lawrennd/nips2014}} and
  blog
  posts\footnote{\url{https://inverseprobability.com/2014/12/16/the-nips-experiment}} all available.
\item
  Reaction triggered by a blog post from Eric Price.\footnote{\url{http://blog.mrtz.org/2014/12/15/the-nips-experiment.html}}
\end{itemize}

Much of the discussion following the conference speculated on the number of consistent accepts in
the process (using the main conference accept rate as a proxy). It
therefore produces numbers that don't match ours above. This is because
the computed accept rate of the individual committees is different from
that of the main conference. This could be due to a bias for the
duplicated papers, or statistical sampling error. To explore the issue, we do some statistics below. First, to get the reader primed for thinking about
these numbers we discuss some context for them.

\subsection{A Random Committee @ 25\% acceptance rate}\label{a-random-committee-25}
The first context we can place around the numbers is what would have
happened at the `random conference' where we simply accept a quarter of
papers at random. In this NeurIPS the expected numbers of accepts would
then have been given as in Table \ref{table-random-committee}.We consider a random committee that has a 25\% accept rate as it makes the numbers easier to remember and it's close to the observed accept rate of 23\%.

\begin{table}[htb]
\caption{Table shows the expected values for the confusion matrix if the committee was making decisions totally at random.}
\label{table-random-committee}
\centering

  \begin{tabular}{lc|c|c|}
  & & \multicolumn{2}{c}{Committee 1} \\
  & & Accept & Reject \\ \hline
  \multirow{2}{*}{Committee 2} & Accept & 10.4 (1 in 16) & 31.1 (3 in 16) \\ 
  & Reject & 31.1 (3 in 16) & 93.4 (9 in 16)
  \end{tabular}
\end{table}

And for this set up we would expect \emph{inconsistency} of 3 in 8
(37.5\%) \emph{accept precision} of 1 in 4 (25\%) and a \emph{reject
precision} of 3 in 4 (75\%) and a \emph{agreed accept rate} of 1 in 10
(10\%). The actual committee made improvements on these numbers, the
accept precision was markedly better with 50\%: twice as many consistent
accept decisions were made than would be expected if the process had
been performed at random and only around two thirds as many inconsistent
decisions were made as would have been expected if decisions were made
at random. However, we should treat all these figures with some
skepticism until we've performed some estimate of the uncertainty
associated with them (see Appendix~\ref{uncertainty-accept-rate}).

\section{Reviewer Calibration}
\label{app:reviewer-calibration}

Calibration of reviewers is the process where different interpretations
of the reviewing scale are addressed. Previous published models used for calibration include the model used in 2006 by John Platt \citep{Platt-calibration12} and the Bayesian variant used in 2013 by Welling and Ghahramani \cite{Ge-bayesian15}. You can find the code recreating our calibration in this Jupyter notebook \url{https://github.com/lawrennd/neurips2014/blob/master/notebooks/Reviewer%20Calibration.ipynb}.

At decision time for NeurIPS 2014, we didn't have access to the \cite{Ge-bayesian15} model and ended up proposing our own approach that turns out to also be a Bayesian interpretation of \cite{Platt-calibration12}. Our assumption is
that the score from the \(j\)th reviewer for the \(i\)th paper is given
by \[
y_{i,j} = f_i + b_j + \epsilon_{i, j},
\] 
where \(f_i\) is the `objective quality' of paper \(i\) and \(b_j\)
is an offset associated with reviewer \(j\). \(\epsilon_{i,j}\) is a
subjective quality estimate which reflects how a specific reviewer's
opinion differs from other reviewers (such differences in opinion may be
due to differing expertise or perspective). The underlying `objective
quality' of the paper is assumed to be the same for all reviewers and
the reviewer offset is assumed to be the same for all papers.

If we have \(n\) papers and \(m\) reviewers, then this implies \(n\) +
\(m\) + \(nm\) values need to be estimated. Naturally this is too many,
and we can start by assuming that the subjective quality is drawn from a
normal density with variance \(\sigma^2\) \[
\epsilon_{i, j} \sim N(0, \sigma^2 \mathbf{I})
\] which reduces us to \(n\) + \(m\) + 1 parameters. Further we can
assume that the objective quality is also normally distributed with mean
\(\mu\) and variance \(\alpha_f\), 
\[
f_i \sim N(\mu, \alpha_f)
\]
this now reduces us to \(m\)+3 parameters. However, we only have
approximately \(4m\) observations (4 papers per reviewer) so parameters
may still not be that well determined (particularly for those reviewers
that have only one paper to assess). We assume that reviewer
offset is normally distributed with zero mean, 
\[
b_j \sim N(0, \alpha_b),
\] 
leaving us only four parameters: \(\mu\), \(\sigma^2\), \(\alpha_f\)
and \(\alpha_b\). Combined together these three assumptions imply that
\[
\mathbf{y} \sim N(\mu \mathbf{1}, \mathbf{K}),
\]
where \(\mathbf{y}\) is a vector of stacked scores \(\mathbf{1}\) is
the vector of ones and the elements of the covariance function are given
by 
\[
k(i,j; k,l) = \delta_{i,k} \alpha_f + \delta_{j,l} \alpha_b + \delta_{i, k}\delta_{j,l} \sigma^2,
\] 
where $\delta_{\cdot,\cdot}$ is Kronecker delta function, \(i\) and \(j\) are the index of first paper and reviewer, and
\(k\) and \(l\) are the index of second paper and reviewer. The mean is
easily estimated by maximum likelihood and is given as the mean of all
scores.

We now reparametrize  to an overall scale
\(\alpha_f\), and normalized variance, \[
k(i,j; k,l) = \alpha_f\left(\delta_{i,k}  + \delta_{j,l} \frac{\alpha_b}{\alpha_f} + \delta_{i, k}\delta_{j,l} \frac{\sigma^2}{\alpha_f}\right)
\]
which we rewrite to give two ratios: offset/signal ratio,
\(\hat{\alpha}_b\) and noise/signal \(\hat{\sigma}^2\) ratio. \[
k(i,j; k,l) = \alpha_f\left(\delta_{i,k}  + \delta_{j,l} \hat{\alpha}_b + \delta_{i, k}\delta_{j,l} \hat{\sigma}^2\right).
\] 
The advantage of this parameterization is it allows us to optimize
\(\alpha_f\) with a fixed-point equation. This leaves us with two free parameters, that we can
explore on a grid. It is in these parameters that we expect the
remaining underdetermindness of the model. We expect \(\alpha_f\) to be
well determined because the negative log likelihood is now 
\[
\frac{|\mathbf{y}|}{2}\log\alpha_f + \frac{1}{2}\log  \left|\hat{\mathbf{K}}\right| + \frac{1}{2\alpha_f}\mathbf{y}^\top \hat{\mathbf{K}}^{-1} \mathbf{y},
\]
where \(|\mathbf{y}|\) is the length of \(\mathbf{y}\) (i.e.~the
number of reviews) and \(\hat{\mathbf{K}}=\alpha_f^{-1}\mathbf{K}\) is
the scale normalized covariance. This negative log likelihood is easily
minimized to recover 
\[
\alpha_f = \frac{1}{|\mathbf{y}|} \mathbf{y}^\top \hat{\mathbf{K}}^{-1} \mathbf{y}.
\] 
A Bayesian analysis of this parameter is possible with gamma priors,
but it would merely show that this parameter is extremely well
determined (the degrees of freedom parameter of the associated
Student-\(t\) marginal likelihood scales will the number of reviews,
which will be around \(|\mathbf{y}| \approx 6,000\) in our case).

So, we propose to proceed as follows. Set the mean from the reviews
(\(\mu\)) and then choose a two-dimensional grid of parameters for
reviewer offset and diversity. For each parameter choice, optimize to
find \(\alpha_f\) and then evaluate the likelihood. Worst case this will
require us inverting \(\hat{\mathbf{K}}\), but if the reviewer paper
groups are disconnected, it can be done a lot quicker.  In practice, we composed the model as a Gaussian process and use of the GPy software for parameter fitting \citep{Gpy-2012}. The full
analysis is described in a blog post.\footnote{\url{https://inverseprobability.com/2014/08/02/reviewer-calibration-for-nips}}.  The parameters of the fitted model are given in the main text in
Table \ref{table-fitted-calibration-parameters}.

\section{Uncertainty: Accept Rate}\label{uncertainty-accept-rate}

To get a handle on the uncertainty around these numbers we'll start by
making use of the
binomial distribution.
First, let's explore the fact that for the overall conference the accept
rate was around 23\%, but for the duplication committees the accept rate
was around 25\%. If we assume decisions are made according to a binomial
distribution, then is the accept rate for the duplicated papers too
high?

To reconstruct these examples or check for any errors in our code you can see our analysis in this Jupyter notebook: \url{https://github.com/lawrennd/neurips2014/blob/master/notebooks/The%20NIPS%20Experiment.ipynb}.

Note that for all our accept probability statistics we used as a
denominator the number of papers that were initially sent for review,
rather than the number where a final decision was made by the program
committee. These numbers are different because some papers are withdrawn
before the program committee makes its decision. Most commonly this
occurs after authors have seen their preliminary reviews: for NeurIPS 2014
we provided preliminary reviews that included paper scores. So for the
official accept probability we use the 170 as denominator. The accept
probabilities were therefore 43 out of 170 papers (25.3\%) for Committee
1 and 44 out of 170 (25.8\%) for Committee 2. This compares with the
overall conference accept rate for papers outside the duplication
process of 349 out of 1508 (23.1\%).

\begin{figure}[htb]
\centering
\includegraphics[width=0.70\textwidth]{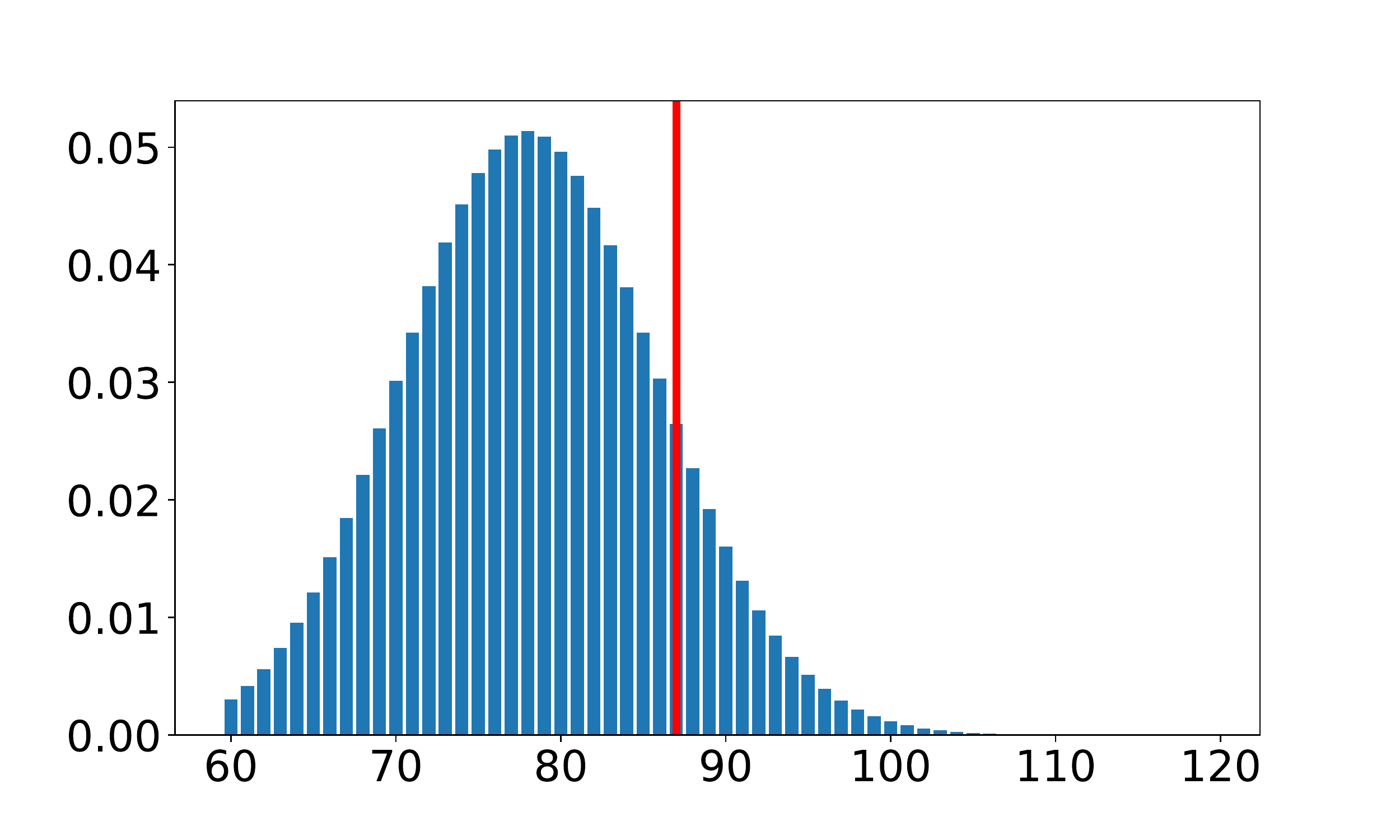}

\caption{Binomial distribution of the number of actual accepted papers we would observe if the true underlying accept probability was $p=0.23$ for all 340 decisions made by the two independent committees involved in the experiment. The observed number of accepted papers is given by a red line that falls well within the probability mass of the binomial.}
\label{fig:uncertainty-accept-rate}
\end{figure}

If the true underlying probability of an accept were 0.23, independent
of the paper, then the probability of generating accepts for any subset
of the papers would be given by a binomial distribution. Combining
across the two committees for the duplicated papers, we see that 87
papers in total were recommended for accept out of a total of 340
possible accepts. The shape of the binomial distribution for these values is 
depicted in Figure~\ref{fig:uncertainty-accept-rate} with the actual number of accepts being shown as a red line.

From the plot, we can see that whilst the accept rate was slightly
higher for duplicated papers it still falls well within the 
probability mass of the binomial distribution implied by a 0.23 accept rate.

Note that Area Chairs knew which papers were duplicates, whereas
reviewers did not. While we stipulated that duplicate papers should not
be any given special treatment, we cannot discount the possibility that
Area Chairs may have given slightly preferential treatment to duplicate
papers.

\subsection{Bayesian Analysis}\label{bayesian-analysis}

Before we start the analysis, it's important to make some statements
about the aims of our modelling here. We will make some simplifying
modelling assumptions for the sake of a model that is understandable. We
are looking to get a handle on the uncertainty associated with some of
the results of the NeurIPS experiment.
Some
preliminary analyses were conducted on blogs.\footnote{See \url{https://inverseprobability.com/2015/01/16/blogs-on-the-nips-experiment} for a summary.} Those
analyses don't have access to information like paper scores etc.. For
that reason we also leave out such information in this preliminary
analysis. We will focus only on the summary results from the experiment:
how many papers were consistently accepted, consistently rejected, or
had inconsistent decisions. For the moment we disregard the information
we have about paper scores.

In our analysis there are three possible outcomes for each paper:
consistent accept, inconsistent decision and consistent reject. So, we
need to perform the analysis with the
multinomial
distribution. The multinomial is parameterized by the probabilities of
the different outcomes. These are our parameters of interest; we would
like to estimate these probabilities alongside their uncertainties. For 
a Bayesian analysis we place a prior density over these
probabilities, then we update the prior with the observed data, that
gives us a posterior density, giving us an uncertainty associated with
these probabilities.

\subsection{Prior Density}\label{prior-density}

For the multinomial likelihood the Dirichlet
density is
conjugate and has
the additional advantage that its parameters can be set to ensure it is
\emph{uninformative}, i.e.~uniform across the domain of the prior.
Combination of a multinomial likelihood and a Dirichlet prior is not
new, and in this domain if we were to consider the mean the posterior
density only, then the approach is known as
Laplace
smoothing.

For our model we are assuming for our prior that the probabilities are
drawn from a Dirichlet as follows, 
\[
p \sim \text{Dir}(\alpha_1, \alpha_2, \alpha_3),
\] 
with \(\alpha_1=\alpha_2=\alpha_3=1\). The Dirichlet density is
conjugate to the
multinomial
distribution, and we associate three different outcomes with the
multinomial. For each of the 166 papers we expect to have a consistent
accept (outcome 1), an inconsistent decision (outcome 2) or a consistent
reject (outcome 3). If the counts four outcome 1, 2 and 3 are
represented by \(k_1\), \(k_2\) and \(k_3\) and the associated
probabilities are given by \(p_1\), \(p_2\) and \(p_3\) then our model
is, \begin{align*}
\mathbf{p}|\boldsymbol{\alpha} \sim \text{Dir}(\boldsymbol{\alpha}) \\
\mathbf{k}|\mathbf{p} \sim \text{mult}(\mathbf{p}).
\end{align*} Due to the conjugacy the posterior is tractable and easily
computed as a Dirichlet (see
e.g.~\cite{Gelman-bayesian13}),
where the parameters of the Dirichlet are given by the original vector
from the Dirichlet prior plus the counts associated with each outcome,
\[
\mathbf{p}|\mathbf{k}, \boldsymbol{\alpha} \sim \text{Dir}(\boldsymbol{\alpha} + \mathbf{k})
\] The mean probability for each outcome is then given by, \[
\bar{p}_i = \frac{\alpha_i+k_i}{\sum_{j=1}^3(\alpha_j + k_j)}.
\] and the variance is \[
\mathrm{Var}[p_i] = \frac{(\alpha_i+k_i) (\alpha_0-\alpha_i + n + k_i)}{(\alpha_0+n)^2 (\alpha_0+n+1)},
\] where \(n\) is the number of trials (166 in our case) and
\(\alpha_0 = \sum_{i=1}^3\alpha_i\). This allows us to compute the
expected value of the probabilities and their variances under the
posterior.

Doing so gives a probability of consistent accept as \(0.136 \pm 0.06\), the
probability of inconsistent decision as \(0.260 \pm 0.09\) and
probability of consistent reject as \(0.60 \pm 0.15\). Recall that if
we'd selected papers at random (with accept rate of 1 in 4) then these
values would have been 1 in 16 (0.0625), 3 in 8 (0.375) and 9 in 16
(0.5625).

\begin{figure}[htb]
\centering
\includegraphics[width=0.90\textwidth]{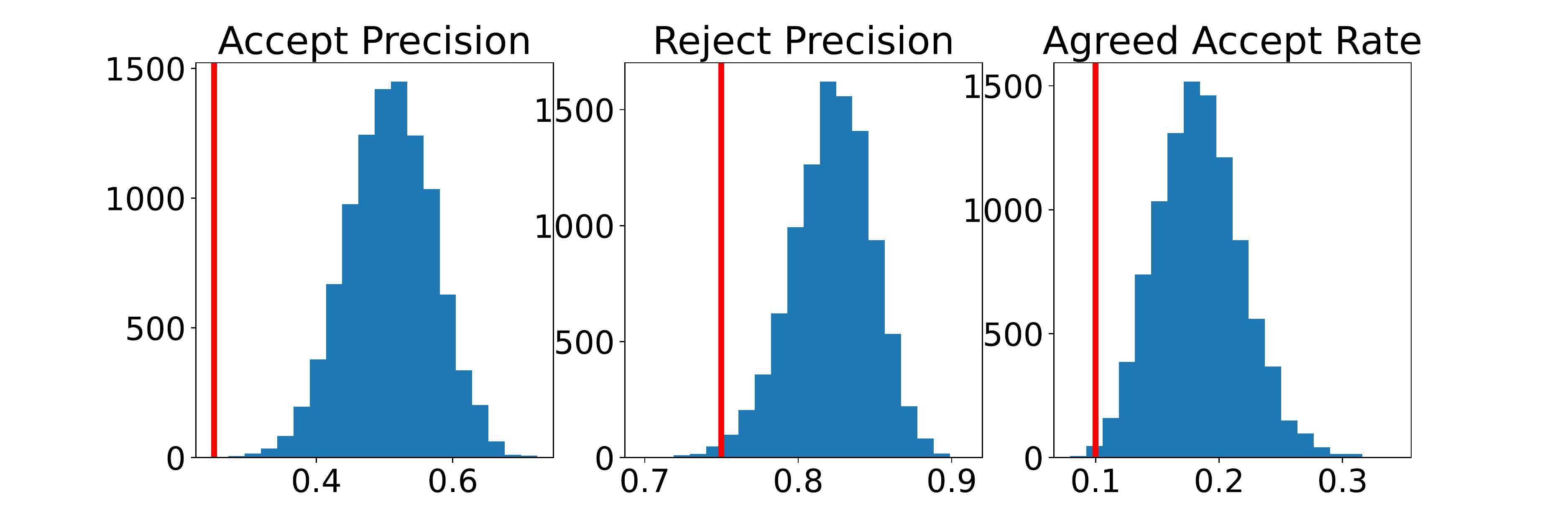}

\caption{Different statistics for the random committee outcomes versus the observed committee outcomes. The red lines indicate results from the NeurIPS experiment, the blue histograms are of Monte Carlo samples from the random committee.}
\label{random-committee-outcomes}
\end{figure}

The other values we are interested in are the accept precision, reject
precision and the agreed accept rate. Computing the probability density
for these statistics is complex: it involves \emph{ratio
distributions}. However, we can use Monte Carlo to estimate the expected
accept precision, reject precision, and agreed accept rate as well as
their variances. We can use these results to give us error bars and
histograms of these statistics giving an accept precision of \(0.51 \pm 0.13\), a reject precision of
\(0.82 \pm 0.05\) and an agreed accept rate of \(0.18 \pm 0.07\). Note
that the `random conference' values of 1 in 4 for accept precision and 3
in 4 for reject decisions are outside the two standard deviation error
bars. If it is preferred medians and percentiles could also be computed
from the samples above, but as we see in histograms  of the results in Figure~\ref{random-committee-outcomes},
the densities look broadly symmetric, so relying on the mean and standard deviation seems appropriate.

From the histograms, it's clear that the observed statistics we see for the committee in the experiment is at the very low end of the distributions generated by Monte Carlo samples from the random committee.

\subsection{Model Choice and Prior
Values}\label{model-choice-and-prior-values}

In the analysis above we've minimised the modeling choices: we made use
of a Bayesian analysis to capture the uncertainty in counts that can be
arising from statistical sampling error. To this end we chose an
uninformative prior over these probabilities. However, one might argue
that the prior should reflect something more about the underlying
experimental structure: for example, we \emph{know} that if the
committees made their decisions independently it is unlikely that we'd
obtain an inconsistency figure much greater than 37.5\% because that
would require committees to explicitly collude to make inconsistent
decisions: the random conference is the worst case. Due to the accept
rate, we also expect a larger number of reject decisions than reject.
This also isn't captured in our prior. Such questions move us into the
realms of modeling the process, rather than performing a sensitivity
analysis. However, if we wish to model the decision process as a whole,
we have a lot more information available, and we should make use of it.
The analysis above is intended to exploit our randomised experiment to
explore how inconsistent we expect two committees to be. It focuses on
that single question; it doesn't attempt to give answers on what the
reasons for that inconsistency are and how it may be reduced. The
additional maths was needed only to give a sense of the uncertainty in
the figures. That uncertainty arises due to the limited number of papers
in the experiment. The simulation in Section~\ref{sec:simulation-of-subjective-scoring} gives an alternative analysis based on the subjective component of the reviewer quality score.

\section{Effect of Late Reviews}
\label{app:effect-of-late-reviews}

\begin{figure}[htb]
\centering
\includegraphics[width=0.70\textwidth]{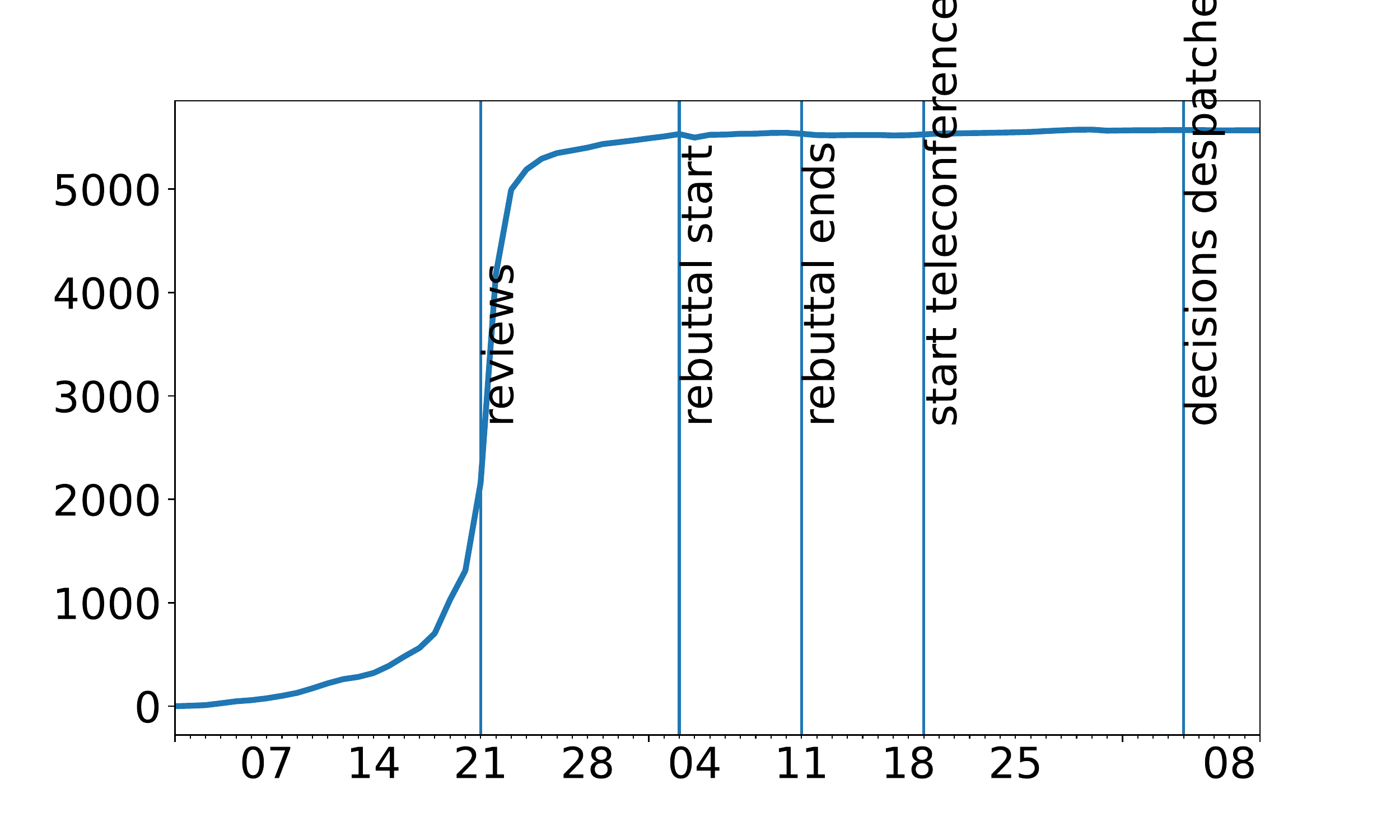}
\caption{Cumulative count of number of received reviews across July and August. Review submission deadline was 21st July.}
\label{review-count}
\end{figure}

Conference reviews do not all come in at the requested time, the cumulative number of reviews that the conference over time is given in Figure \ref{review-count}.

\begin{figure}[htb]
\centering
\includegraphics[width=0.70\textwidth]{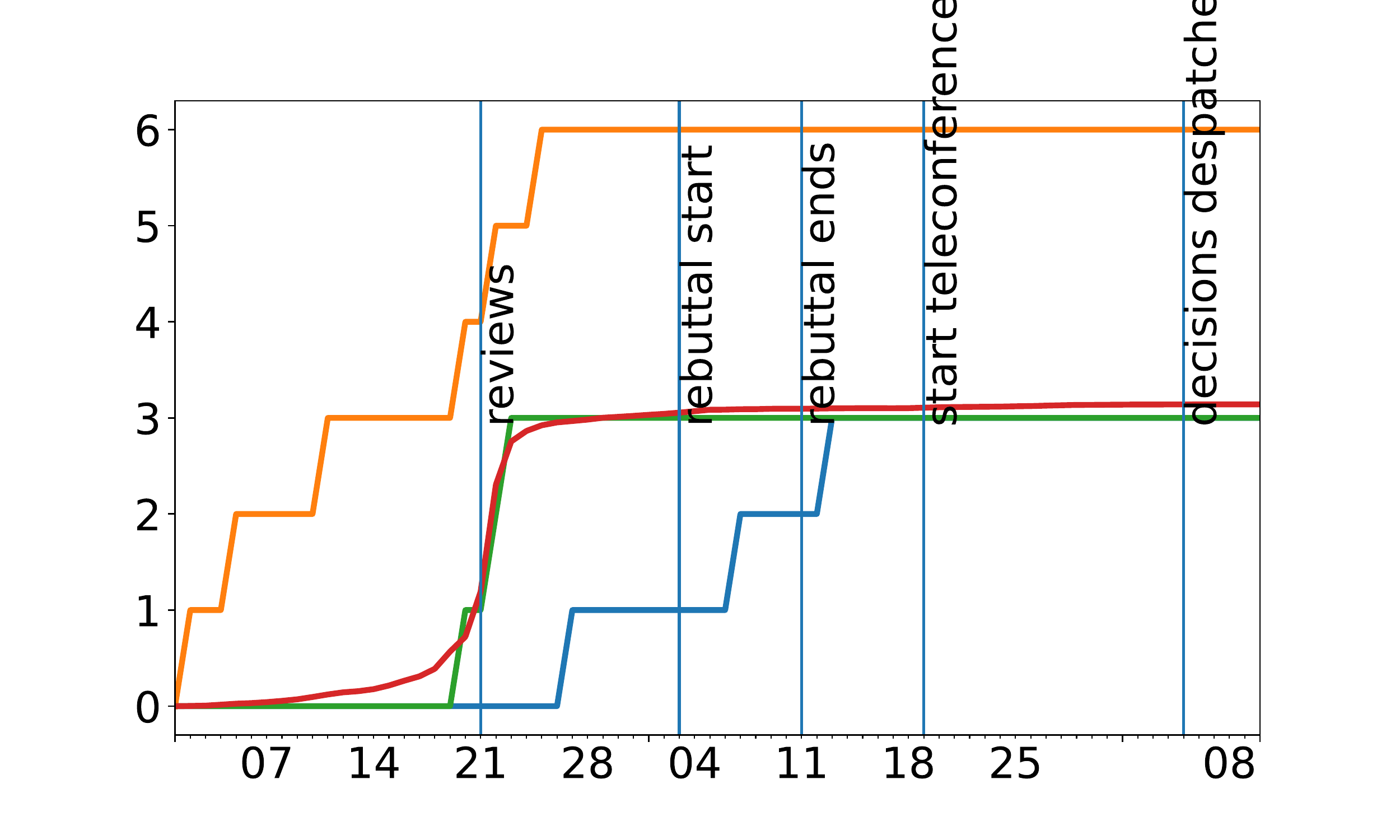}

\caption{Plot representing number of reviewers per paper over time showing maximum number of reviewers per paper, minimum, median, and mean. }
\label{number-of-reviews-over-time}
\end{figure}

The Program Committee worked hard to ensure that all  papers had three reviews
before the start of the rebuttal. In this section we explore differences between late submitted reviews and those submitted on time. Code for recreating this analysis is available in this Jupyter notebook: \url{https://github.com/lawrennd/neurips2014/blob/master/notebooks/NIPS%202014%20Late%20Review%20Observations.ipynb}.  In Figure \ref{number-of-reviews-over-time}  we show the  overall statistics of what the count of reviewers per
paper were, we plot mean, maximum, median, and minimum over time. 

\begin{figure}[htb]
\centering
\includegraphics[width=0.70\textwidth]{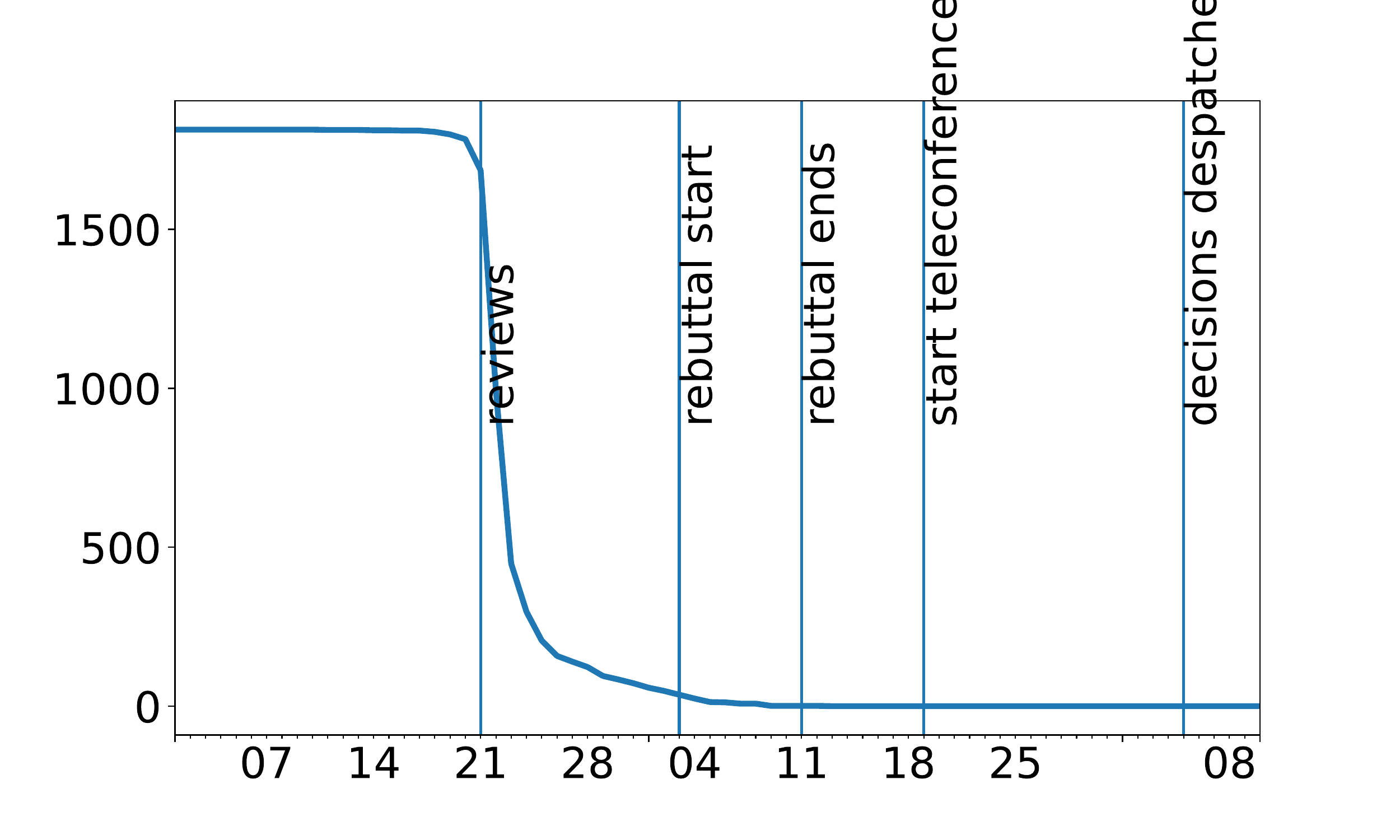}

\caption{Number of papers with less than three reviewers as a function of time.}
\label{paper-short-reviews}
\end{figure}

In Figure~\ref{paper-short-reviews} we show the number of
papers that had less than three reviews across the review period.

\begin{figure}[htb]
\centering
\includegraphics[width=0.70\textwidth]{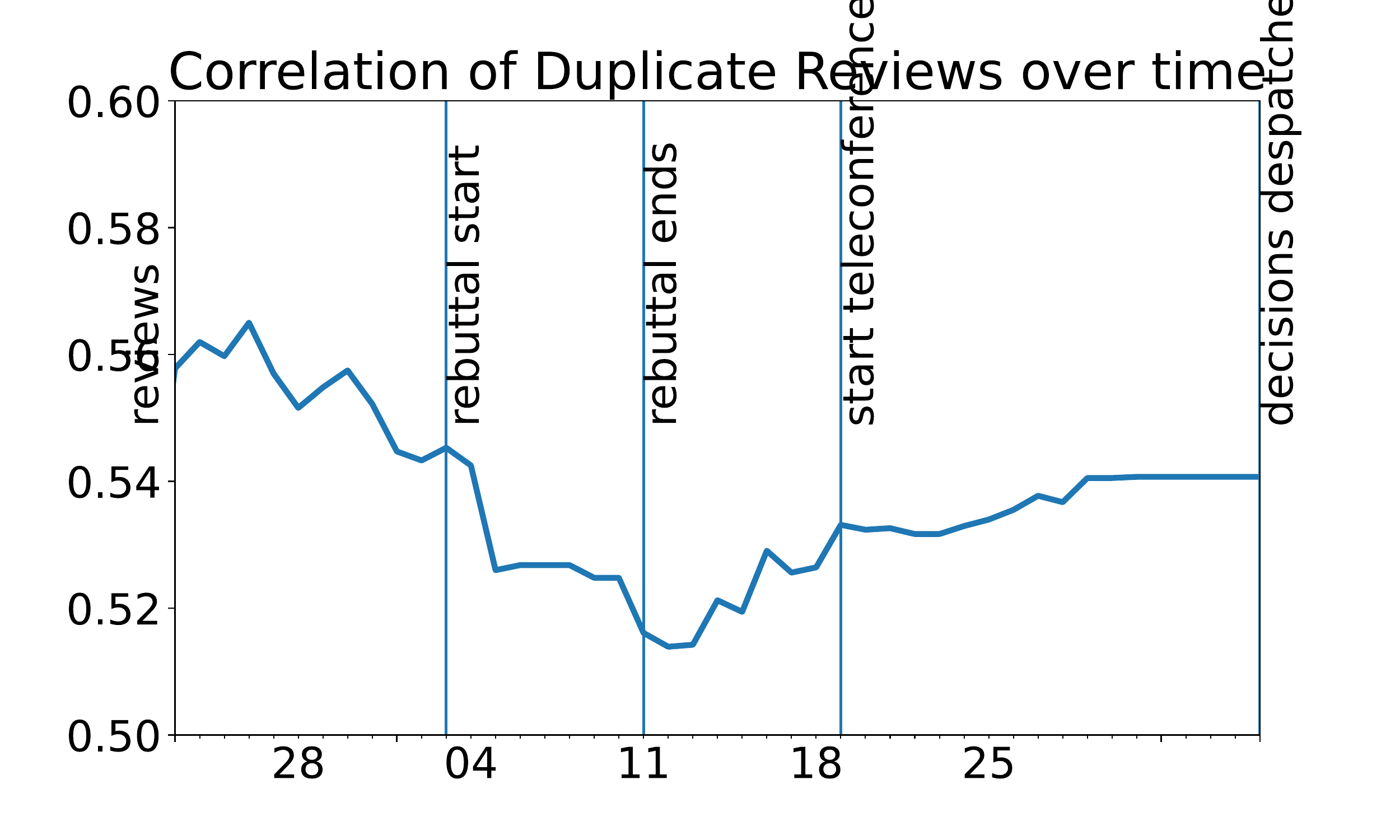}

\caption{Average correlation of duplicate papers over time.}
\label{correlation-duplicate-reviews}
\end{figure}

However, while observing the late reviewer scores coming in, the correlation between the two committees reduced (see Figure~\ref{correlation-duplicate-reviews}). Correlation starts high, but it drops as reviewers were chased to submit, then it recovers during the reviewer discussion period. This trend triggered a question: are late reviewers damaging the reviewing process?

\begin{figure}[htb]
\centering
\includegraphics[width=0.70\textwidth]{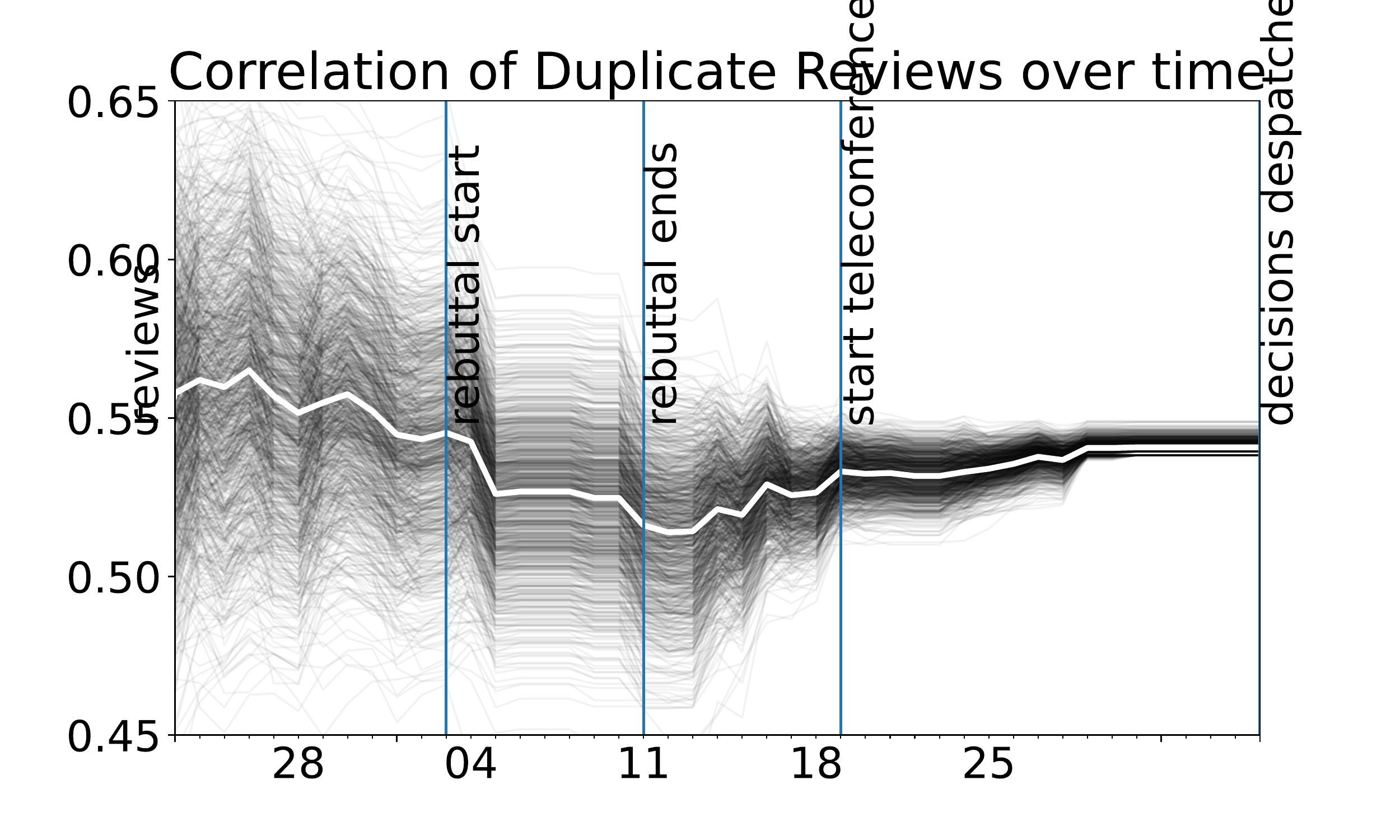}

\caption{Average correlation of duplicate papers over time. To give an estimate of the uncertainty the correlation is computed with bootstrap samples. Here to allow comparison between the trend lines similar, the bootstrap samples are set so they converge on the same point on the right of the graph.}
\label{correlation-duplicate-reviews-bootstrap}
\end{figure}

These changes in correlation were point estimates based on a relatively low number of papers. To get a sense of the uncertainty overtime, we computed bootstrap estimates of the correlation over time. These estimates are shown in Figure~ \ref{correlation-duplicate-reviews-bootstrap}.

The bootstrap estimates show how much uncertainty there is in these correlation estimates, and caution us about how much credit to place against any interpretation of these correlations over time, but with curiosity already piqued by the original plot, we decided to compare the quality of reviews between those that were on-time and those that were late.

\subsection{Review Confidence}\label{review-confidence}

\begin{figure}[htb]
\centering
\includegraphics[width=0.70\textwidth]{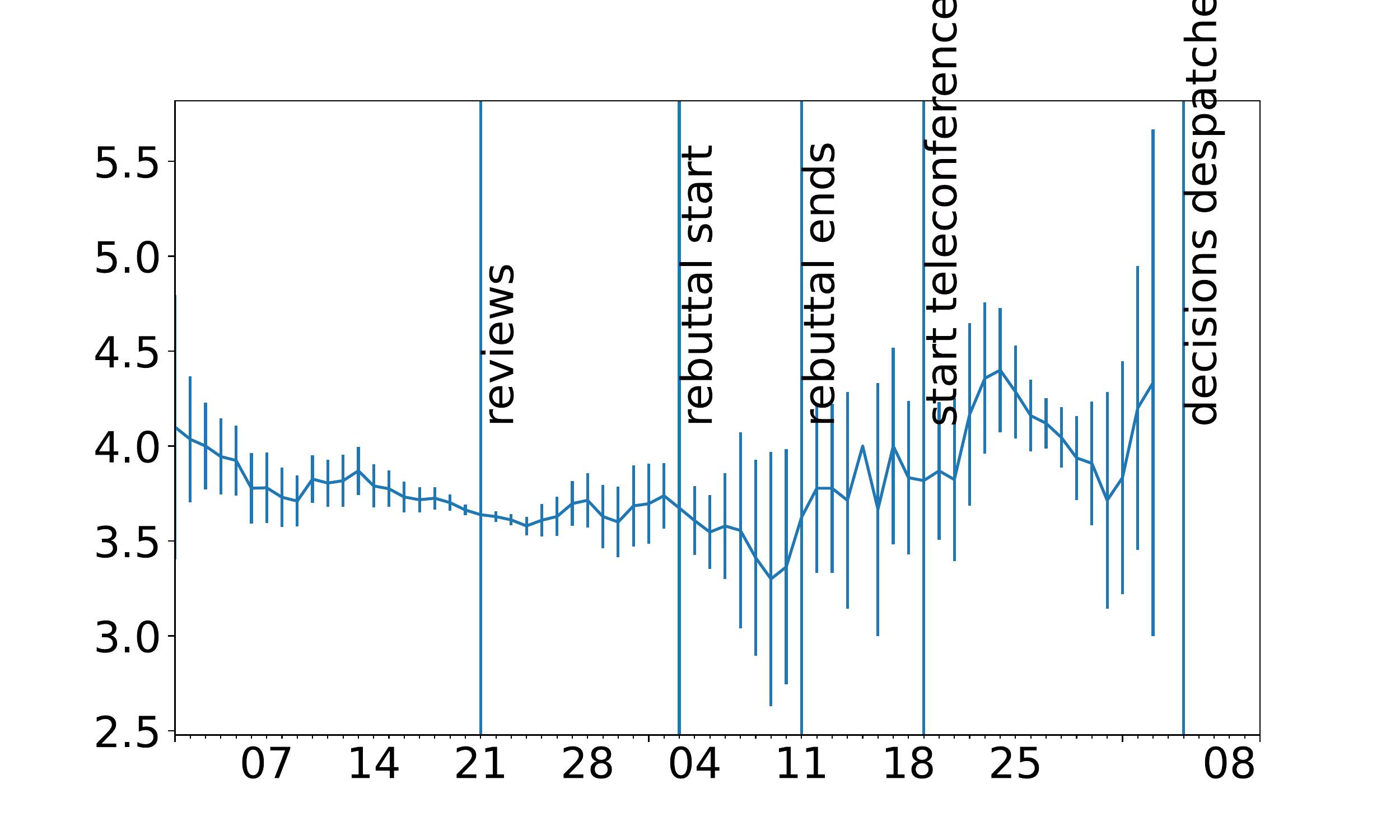}

\caption{Average confidence of reviews as computed across a four-day moving window, plot includes standard error the mean estimate.}
\label{review-confidence-time}
\end{figure}

First, we consider whether the confidence of reviewes varied over time. We considered a moving four day window across the review submission period and in Figure~\ref{review-confidence-time} we show the mean review confidence over time, alongside 95\% confident intervals computed from the standard errors for the mean estimate. 

Again, a definitive trend is difficult to pick out, but it's plausible that there's a reduction in confidence as we pass the review deadline on 21st July, so the next step was to explore whether that's a statistically significant reduction.

\begin{figure}[htb]
\centering
\includegraphics[width=0.50\textwidth]{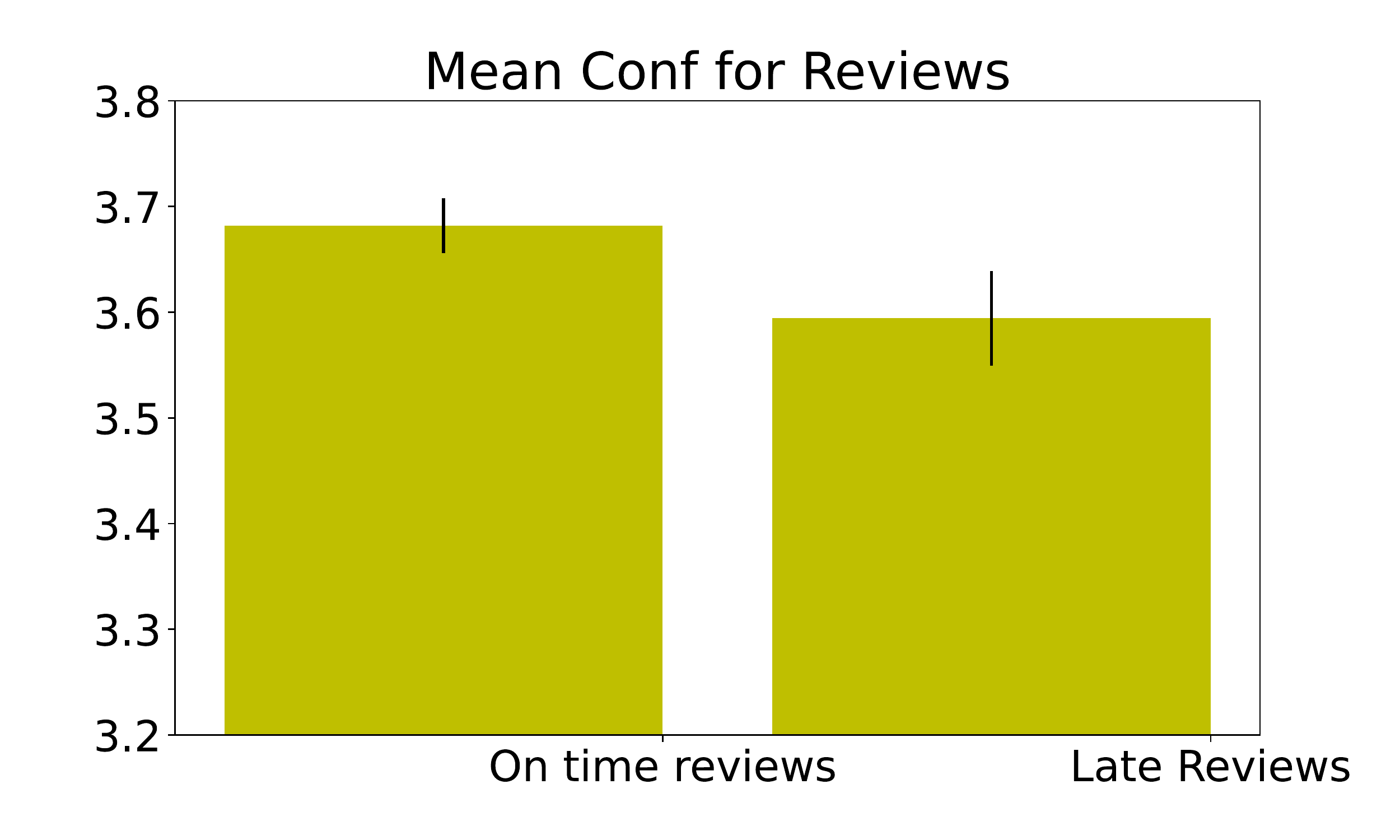}

\caption{Average confidence for reviews that arrived
before 21st July (the reviewing deadline) and reviews that arrived
after. Histogram shows mean values and confidence intervals. A
\(t\)-test shows the difference to be significant with a \(p\)-value of
0.048\%, although the magnitude of the difference is small (about
0.1).} \label{review-confidence-early-late}
\end{figure}

We looked at the average confidence for
reviews that arrived before 21st July (the reviewing deadline) and
reviews that arrived after the 21st July (i.e.~those that were chased or
were allocated late) but before the rebuttal period started (4th
August). In Figure~\ref{review-confidence-early-late} we show the average estimate for these two groups along with error bars from standard error.

Viewing the plot suggests that there's a small but significant difference between
the average confidence of the submitted reviews before and after the
deadline, the statistical significance is confirmed with a \(t\)-test
with a \(p\)-value at 0.048\%. The magnitude of the difference is small
(about 0.1) but may indicate a tendency for later reviewers to be a
little more rushed.

\subsection{Quality and Impact Scores}\label{quality-score}

\begin{figure}[htb]
\centering
\includegraphics[width=0.70\textwidth]{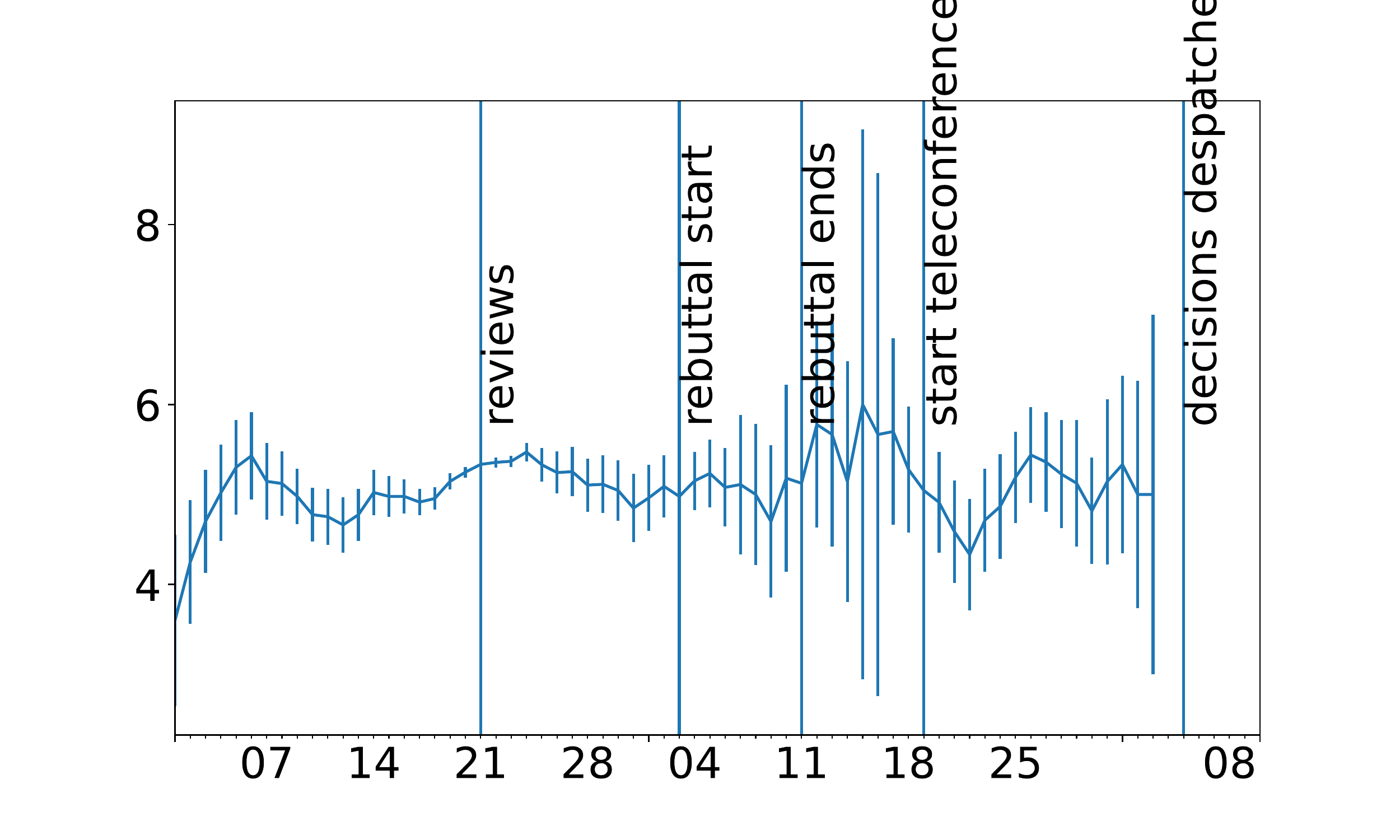}

\caption{Plot of average review quality score as a function of time using a four day moving window. Standard error is also shown in the plot.}
\label{review-quality-time}
\end{figure}

\begin{figure}[htb]
\centering
\includegraphics[width=0.50\textwidth]{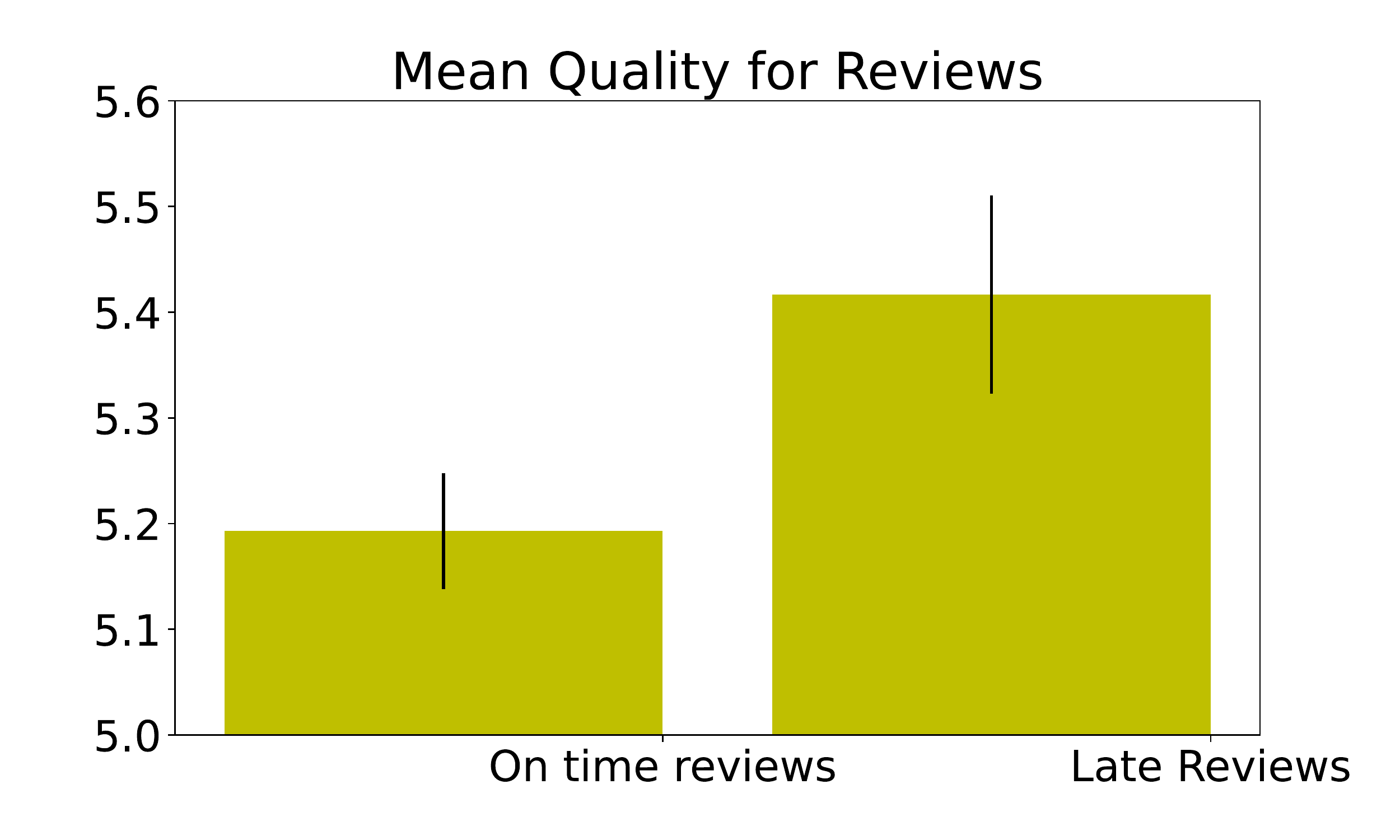}

\caption{Bar plot of average quality scores for on-time
reviews and late reviews, standard errors shown. Under a \(t\)-test the
difference in values is statistically significant with a \(p\)-value of
0.007\%.} \label{review-quality-early-late}
\end{figure}

Following this analysis around confidence, we performed the same analysis for both the quality scores and the impact scores. The results are shown in Figures
\ref{review-quality-time} and \ref{review-quality-early-late}.

There is another statistically significant difference between perceived
quality scores after the reviewing deadline than before. On average
reviewers tend to be more generous in their quality perceptions when the
review is late. The \(p\)-value is computed as 0.007\%. We can also
check if there is a similar on the impact score (see Section \ref{paper-scoring-and-reviewer-instructions} for more details on the scores). The impact score was
introduced by Ghahramani and Welling in 2013 to get reviewers not just to think
about the technical side of the paper, but whether it is driving the
field forward. The score is binary, with 1 being for a paper that is
unlikely to have high impact and 2 being for a paper that is likely to
have a high impact.

\begin{figure}[htb]
\centering
\includegraphics[width=0.70\textwidth]{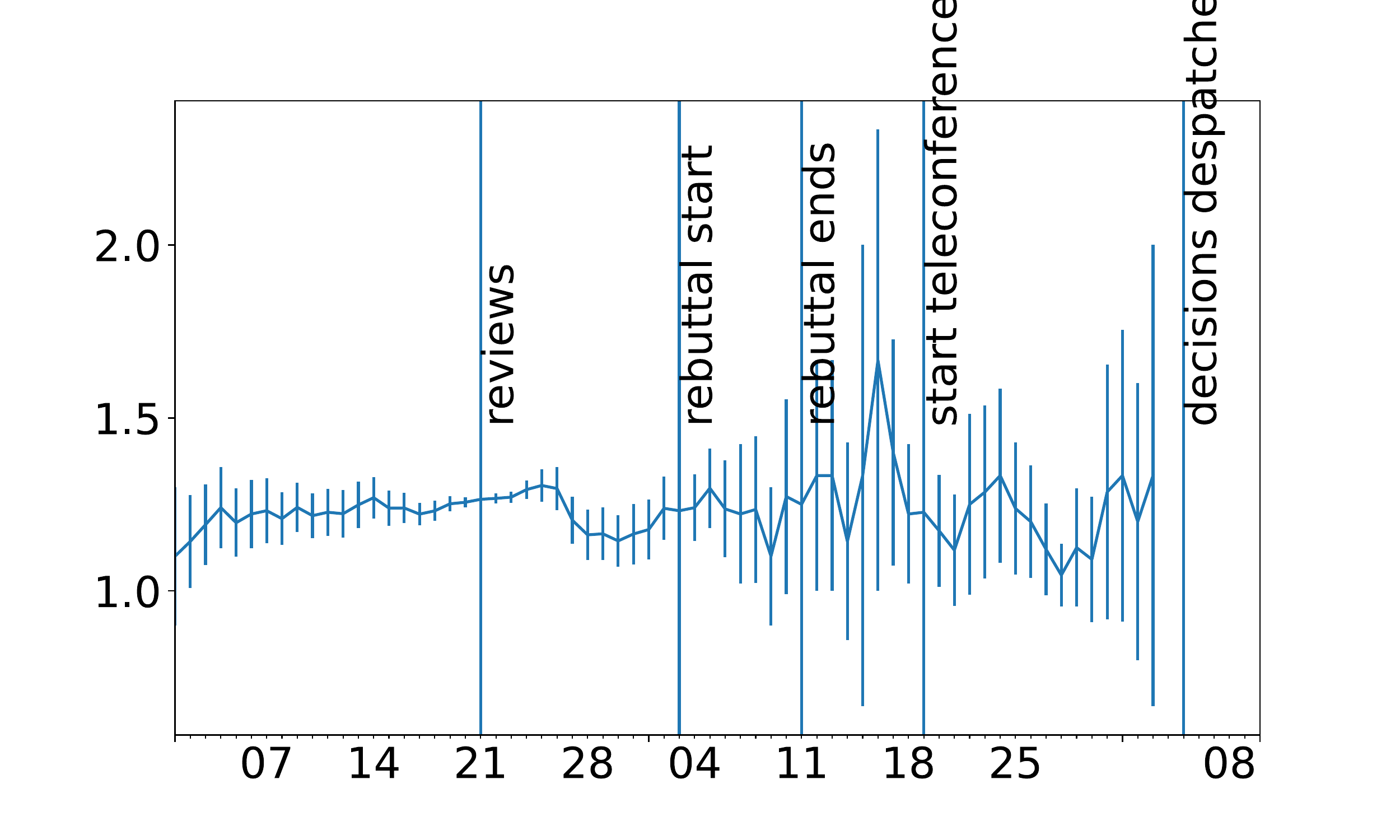}

\caption{Average impact score for papers over time, again using a moving average with a window of four days and with standard error of the mean computation shown.}
\label{review-impact-time}
\end{figure}

\begin{figure}[htb]
\centering
\includegraphics[width=0.50\textwidth]{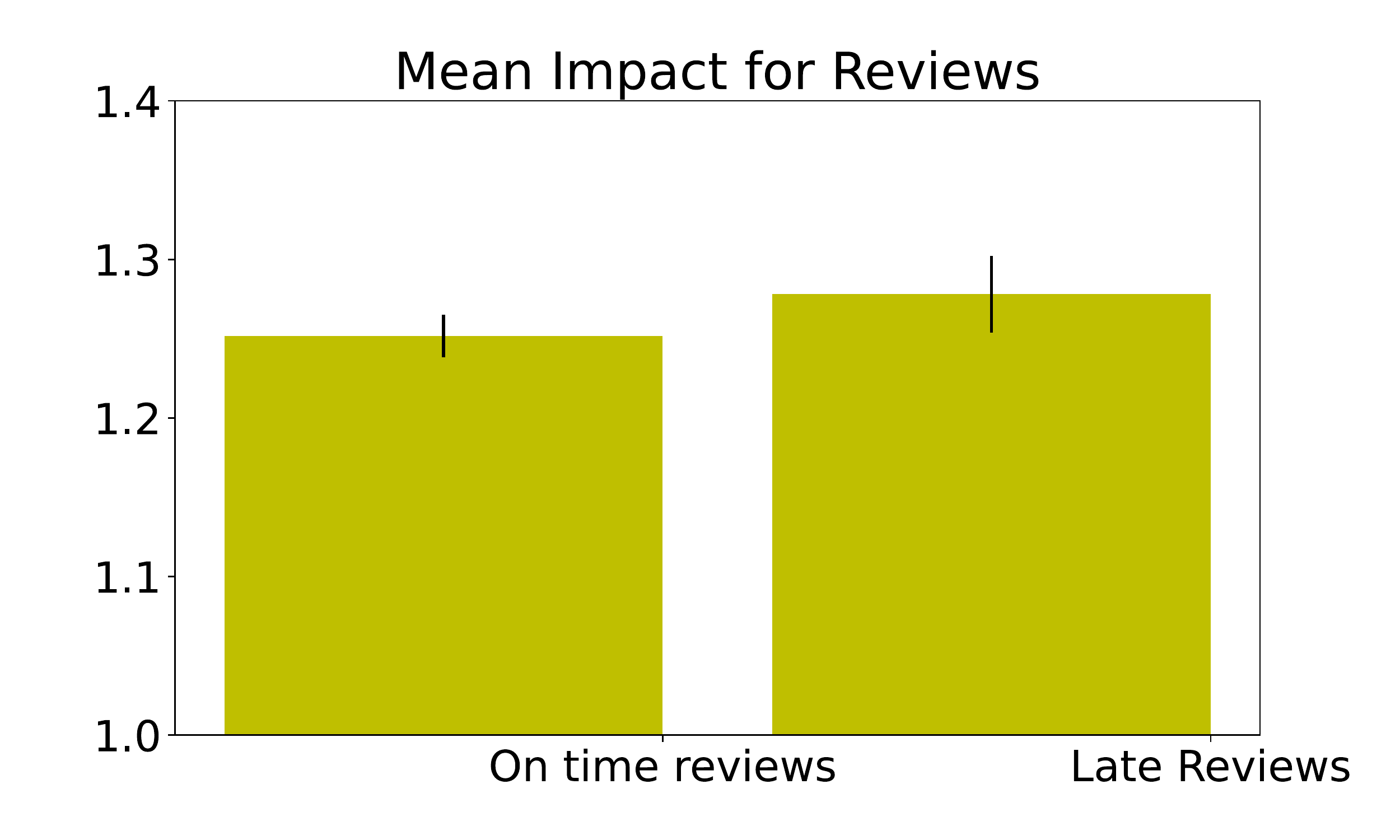}

\caption{Bar plot showing the average impact score of
reviews submitted before the deadline and after the deadline. The
difference in means did not prove to be statistically significant under
a \(t\)-test (\(p\)-value 5.9\%).} \label{review-impact-early-late}
\end{figure}

We find the difference is not quite statistically significant for the
impact score (\(p\)-value of 5.9\%), but if anything, there is a trend
to have slightly higher impacts for later reviews (see Figures
\ref{review-impact-time} and \ref{review-impact-early-late}).

\subsection{Review Length}\label{review-length}

\begin{figure}[htb]
\centering
\includegraphics[width=0.70\textwidth]{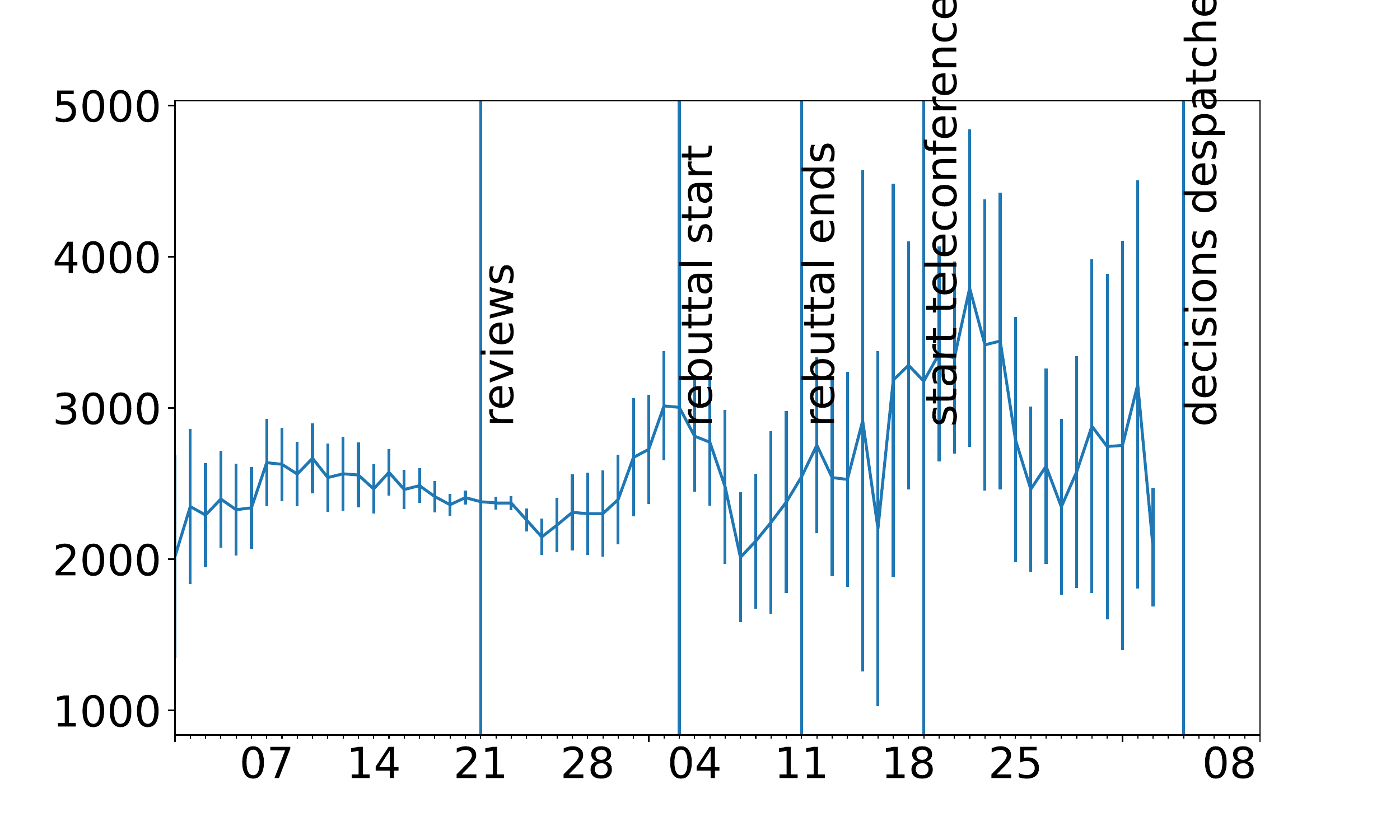}

\caption{Average length of reviews submitted plotted as a function of time with standard error of the mean computation included.}
\label{review-length-time}
\end{figure}

\begin{figure}[htb]
\centering
\includegraphics[width=0.50\textwidth]{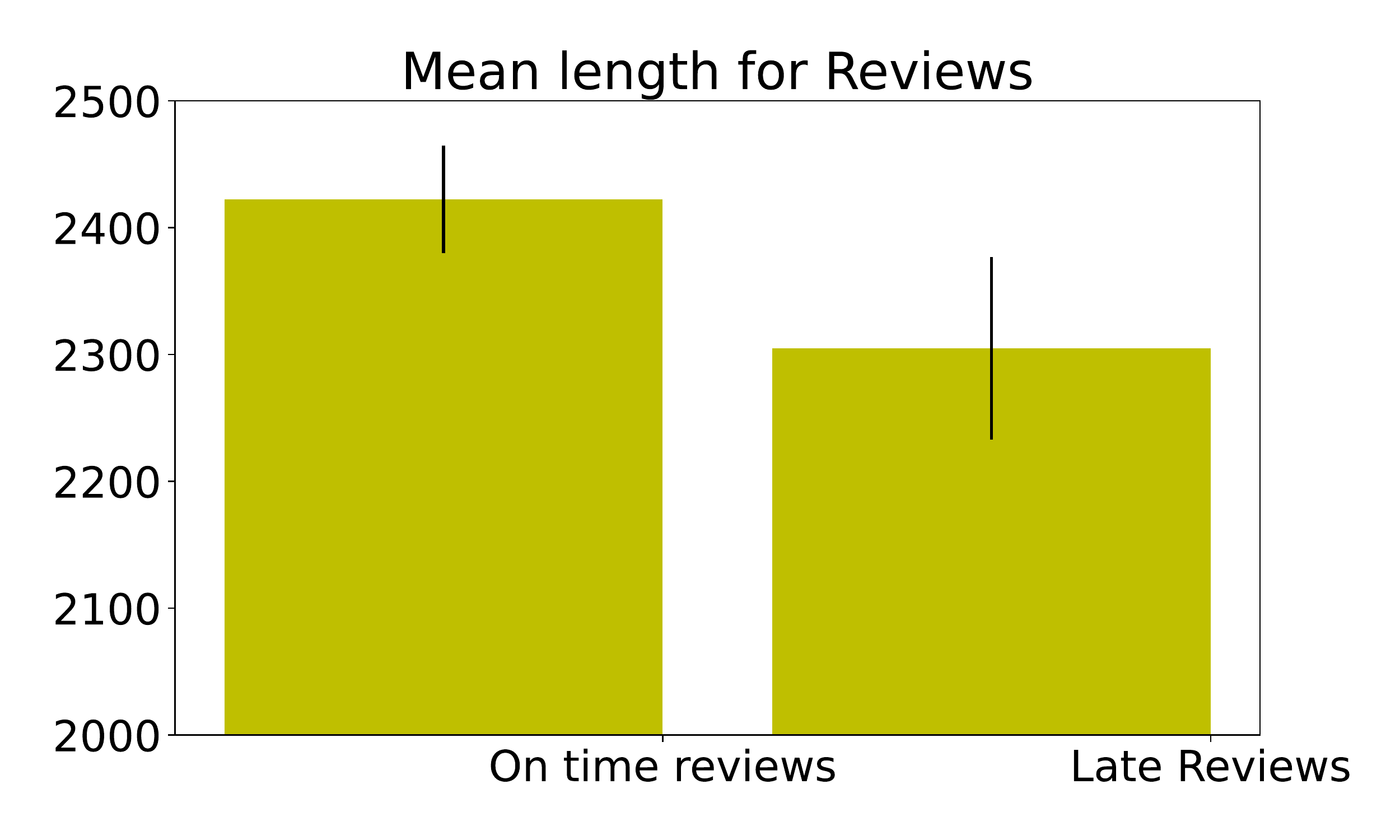}

\caption{Bar plot of the average length of reviews
submitted before and after the deadline with standard errors included.
The difference of around 100 words is statistically significant under a
\(t\)-test (\(p\)-value 0.55\%).} \label{review-length-early-late}
\end{figure}

A final potential indicator of the review quality is the length of the
reviews, we can check if there is a difference between the combined
length of the review summary and the main body comments for late and
early reviews (see Figures \ref{review-length-time} and
\ref{review-length-early-late}).

Once again we find a small but statistically significant difference,
here, as we might expect late reviews are shorter than those submitted
on time, by about 100 words in a 2,400 word review.

\subsection{Late Reviewers Summary}\label{late-reviewers-summary}

In summary we find that late reviews are on average less confident and
shorter, but rate papers as higher quality and perhaps as higher impact.
Each of the effects is small (around 5\%) but overall a picture emerges
of a different category of review arriving from those that delay their
assessment.

\section{Correlation of Quality Scores and
Citation}\label{correlation-of-quality-scores-and-citation}

To revisit the NeurIPS experiment, we traced the fate of accepted papers and all the rejected papers that we could find through searching on Semantic Scholar. This allowed us to associate each paper with a citation count. In this section we explore correlation between those citation counts and the scores that reviewers gave the papers back in 2014.

Code that traces the fate of rejected papers (in terms of their final publication venue) can be found in this Jupyter notebook: \url{https://github.com/lawrennd/neurips2014/blob/master/notebooks/where-do-rejected-papers-go.ipynb} and code that computes the citation scores and compares them to quality scores can be found in this Jupyter notebook: \url{https://github.com/lawrennd/neurips2014/blob/master/notebooks/Measuring%20Impact%20of%20Papers%20Using%20Semantic%20Scholar.ipynb}.

\begin{figure}[htb]
\centering
\includegraphics[width=0.70\textwidth]{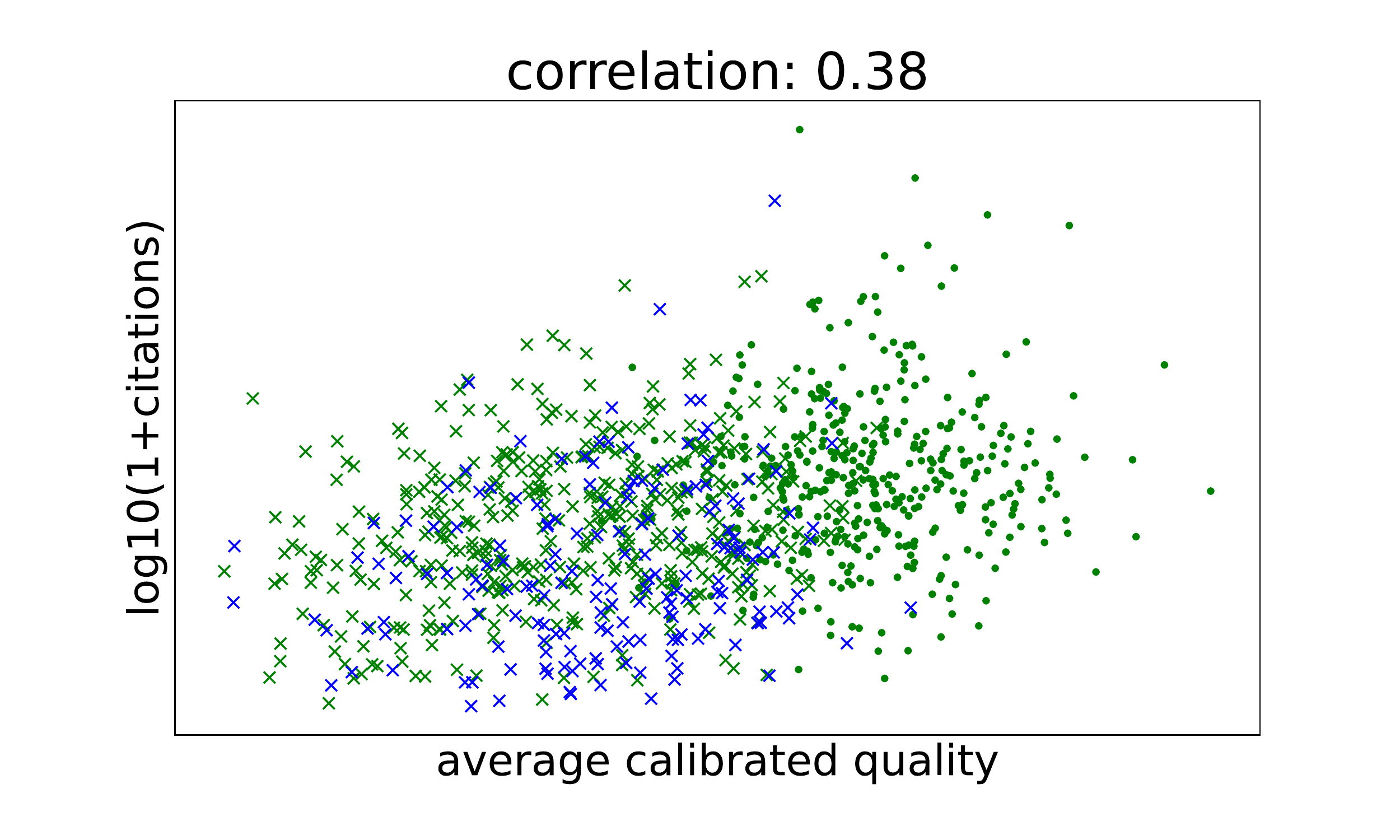}

\caption{Scatter plot of $\log_{10}(1+\text{citations})$ against the average calibrated quality score for all papers. To prevent reidentification of individual papers quality scores and citation count, each point is corrupted by differentially private noise in the plot (correlation is computed before adding differentially private noise). Rejected papers are given as crosses, accepted papers are shown as dots. Papers that found a venue are shown in green. Any paper that is only available through pre-print servers or as an unpublished PDF is shown in blue.}
\label{figure-citations-vs-average-calibrated-quality-all}
\end{figure}

First of all, we plot the correlation between average quality score and the citation count for all papers where we could trace their fate (accepted and rejected). This is shown in  Figure~\ref{figure-citations-vs-average-calibrated-quality-all}. Rejected papers are given as crosses,
accepted papers are given as dots. We include all papers, whether
published in a venue or just available through ArXiv or other preprint
servers. We show the published/non-published quality scores and
\(\log_{10}(1+\text{citations})\) for all papers in the plot below. In
the plot we are showing each point corrupted by some Laplacian noise and
also removing axes. The idea is to give a sense of the distribution
rather than reveal the score of a particular paper.

The correlation between the reviewer scores and the citation impact score looks strong, but there is a counfounder here because we've bundled together papers that were accepted and those that were rejected from the conference. Papers accepted by the conference will have been published earlier and presentation at the conference is likely to have given a  lift to the papers' numbers of citations. So, we analyze these two groups separately.

Looking at the accepted papers only shows a very different picture. As outlined in the main text, 
there is no statistically significant correlation between accepted papers' quality scores
and the number of citations they receive (see Figure \ref{figure-citations-vs-average-calibrated-quality-accept}). 

Conversely, looking at rejected papers only, we do see a correlation (Figure~\ref{figure-citations-vs-average-calibrated-quality-reject}),
with higher scoring papers achieving more citations on average. This,
combined with the lower average number of citations in the rejected
paper group, alongside their lower average scores, explains the
correlation we observed when analyzing the two groups of papers together.

Welling and Ghahramani introduced an ``impact'' score in NeurIPS 2013,
we might expect the impact score to show correlation. And indeed,
despite the lower range of the score (a reviewer can score either 1 or
2) we do see \emph{some} correlation, although it is relatively weak (Figure \ref{figure-citations-vs-average-impact-accept}).

Finally, we also looked at correlation between the \emph{confidence}
score and the citation count. Here correlation is somewhat stronger (0.25) and is statistically significant (see Figure~\ref{figure-citations-vs-average-confidence-accept}). Why should
confidence be an indicator of higher citations? A plausible explanation
is that there is confounder driving both variables. For example, it
might be that papers which are easier to understand (due to elegance of
the idea, or quality of exposition) inspire greater reviewer confidence
and increase the number of citations.


\end{document}